\documentclass[preprint,a4paper,10pt,3p]{elsarticle}

\usepackage{amssymb}

\newdefinition{rmk}{Remark}

\usepackage{physics}

\usepackage{siunitx}
\usepackage{xcolor}
\usepackage{psfrag}
\usepackage{booktabs}
\usepackage{subfigure}

\journal{arXiv}

\begin{document}

\begin{frontmatter}


\title{A Versatile SPH Modeling Framework for Coupled Microfluid-Powder Dynamics in Additive Manufacturing: Binder Jetting, Material Jetting, Directed Energy Deposition and Powder Bed Fusion}

\author[1,2]{Sebastian L. Fuchs}
\ead{sebastian.fuchs@tum.de}

\author[1]{Patrick M. Praegla}
\ead{patrick.praegla@tum.de}

\author[2,3]{Christian J. Cyron}
\ead{christian.cyron@tuhh.de}

\author[1]{Wolfgang A. Wall}
\ead{wolfgang.a.wall@tum.de}

\author[1]{Christoph Meier\corref{cor}}
\ead{christoph.anton.meier@tum.de}

\cortext[cor]{corresponding author}

\address[1]{Institute for Computational Mechanics, Technical University of Munich, Garching, Germany}

\address[2]{Institute for Continuum and Material Mechanics, Hamburg University of Technology, Hamburg, Germany}

\address[3]{Institute of Material Systems Modeling, Helmholtz-Zentrum Hereon, Geesthacht, Germany}

\begin{abstract}
Many additive manufacturing (AM) technologies rely on powder feedstock, which is fused to form the final part either by melting or by chemical binding with subsequent sintering. In both cases, process stability and resulting part quality depend on dynamic interactions between powder particles and a fluid phase, i.e., molten metal or liquid binder. The present work proposes a versatile computational modeling framework for simulating such coupled microfluid-powder dynamics problems involving thermo-capillary flow and reversible phase transitions. In particular, a liquid and a gas phase are interacting with a solid phase that consists of a substrate and mobile powder particles while simultaneously considering temperature-dependent surface tension and wetting effects. In case of laser-metal interactions, the effect of rapid evaporation is incorporated through additional mechanical and thermal interface fluxes. All phase domains are spatially discretized using smoothed particle hydrodynamics. The method's Lagrangian nature is beneficial in the context of dynamically changing interface topologies. Special care is taken in the formulation of phase transitions, which is crucial for the robustness of the computational scheme. While the underlying model equations are of a very general nature, the proposed framework is especially suitable for the mesoscale modeling of various AM processes. To this end, the generality and robustness of the computational modeling framework is demonstrated by several application-motivated examples representing the specific AM processes binder jetting, material jetting, directed energy deposition, and powder bed fusion. Among others, it is shown how the dynamic impact of droplets in binder jetting or the evaporation-induced recoil pressure in powder bed fusion leads to powder motion, distortion of the powder packing structure, and powder particle ejection.
\end{abstract}


\end{frontmatter}

\section{Introduction} \label{sec:intro}

\begin{figure}[htbp]
\centering
\includegraphics[width=1.0\textwidth]{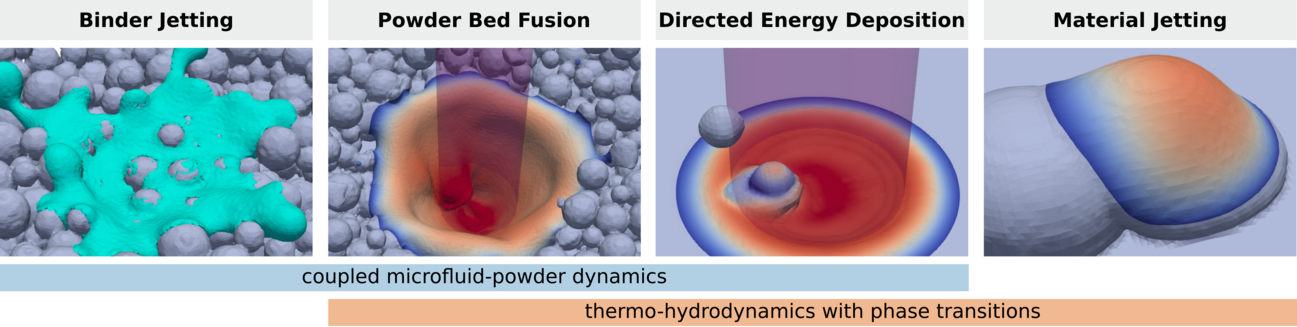}
\caption{Overview of the considered AM processes with classification of two characteristic physical phenomena.}
\label{fig:intro_processes_overview}
\end{figure}

Additive manufacturing (AM) offers great opportunities in product design and manufacturing, and thus received considerable attention in research and industry over the past years. Especially, computational simulation of AM processes helps to gain a deeper understanding of involved process physics. In this spirit, this work proposes a general modeling framework for solving coupled microfluid-powder dynamics problems involving thermo-capillary flow and reversible phase transitions. This modeling framework is especially suitable to examine complex physical phenomena in AM processes such as binder jetting (BJT), material jetting (MJT), directed energy deposition (DED), and powder bed fusion (PBF). In the following, a brief overview of these AM processes is given. Within this scope, two characteristic physical phenomena are identified which play a crucial role, coupled microfluid-powder dynamics (relevant for BJT, PBF, and DED) and thermo-hydrodynamics with phase transitions (relevant for PBF, DED, and MJT), as illustrated in Figure~\ref{fig:intro_processes_overview}.

In BJT, liquid binder droplets ($\sim\SI{80}{\micro\meter}$ diameter) are deposited from a print head onto a previously spread powder layer to bind powder particles together, forming the cross-section of the desired part. Subsequently, a new powder layer is spread (similarly to PBF processes) and the droplet deposition is repeated. The resulting so-called green part is solely held together by the fully solidified binder and requires post-processing to increase mechanical strength, e.g., sintering in case of metals or ceramics~\cite{Gibson2021,Ziaee2019, Mostafaei2021}. Besides challenges in powder handling and spreading as well as post-processing of the part, there is need for a more detailed understanding of the coupled microfluid-powder dynamics during binder jetting, which may significantly affect the packing structure of the powder bed after droplet impact and, eventually, the quality of the final part~\cite{Ziaee2019}. In MJT no additional binder material is required. Instead, the design material, e.g., metal, is processed in liquid state and deposited from a print head. The ejected melt droplets solidify on top of previously deposited material and, thus, layer-by-layer create the final part~\cite{Gibson2021}. Critical aspects of the process are, e.g., the degree of droplet-substrate adhesion, the phenomenon of droplet-droplet coalescence, as well as the rate of cooling and solidification~\cite{Gilani2021}. In DED a focused heat source (laser or electron beam) is used to melt a continuous stream of feedstock material (powder or wire), again to build a part layer-by-layer. A common type of DED process is laser powder deposition (LPD), where a moving deposition head combines the laser heat source and the powder delivery. Through one or more nozzles the deposition head creates a stream of powder particles that is focused at the laser interaction zone. Depending on the processing conditions, the powder either melts during flight or when entering the melt pool below the laser. The shape of the final part is defined by a complex three-dimensional relative motion between substrate and deposition head, which is achieved by moving either the deposition head, or the substrate, or a combination of both~\cite{Gibson2021}. A comprehensive overview of DED is given, e.g., in~\cite{Thompson2015, Shamsaei2015,Svetlizky2021}. In PBF processes, such as selective laser melting (SLM) or electron beam melting (EBM), complex structural components are created by selectively melting and fusing thin layers of metal powder in a layer-by-layer fashion~\cite{Markl2016,Meier2017b,Bayat2021}. These processes are driven by temperature-dependent surface tension, wetting, and capillary forces, which crucially affect the shape of the melt pool~\cite{Korner2011,Korner2013,Meier2017b} and the surface topology of the solidified track. Typically, the peak temperatures in PBF, but also in DED, exceeds the boiling temperature of the liquid metal. Thus, strong evaporation occurs under typical PBF process conditions, giving rise for evaporation-induced recoil pressure, which is the origin for the characteristic keyhole-shaped depression of the melt pool surface, as well as dynamic vapor and gas flows in the build chamber. From experiments it is well-known that the evaporation-induced reoil pressure and vapor/gas flow can lead to powder particle ejection and entrainment within a domain that exceeds the dimensions of the melt pool by far~\cite{Khairallah2016,Liu2016,Gusarov2007,Ly2017,Matthews2016}.

Altogether, the aforementioned AM processes open up great opportunities in product design and manufacturing. However, many of the underlying physical phenomena as well as their influence on process stability and part quality are still insufficiently understood. In particular, this involves the microfluidic behavior of liquid binder or molten metal, especially when considering the coupling with powder dynamics and/or temperature-induced phase transition phenomena. Physics-based modeling and predictive simulation has the potential to gain a better understanding of the governing process physics and the optimal processing conditions to improve, e.g., processing hardware, strategies, and materials, and to mitigate process instabilities and the creation of defects~\cite{Meier2021b}.

Recent modeling approaches of BJT processes have focused on the microfluid dynamics of liquid binder droplets infiltrating the powder bed, e.g., to determine equilibrium saturations~\cite{Miyanaji2018}, or to predict droplet impact and penetration dynamics~\cite{Tan2016,Deng2022}. In the field of MJT, several modeling approaches, e.g., based on the finite difference method, the finite volume method, or the finite element method, have been proposed~\cite{Pasandideh-Fard2002,Li2012,Gilani2021} to study the thermo-hydrodynamics of droplet impact and solidification. An extensive review of computational models for the simulation of DED processes, which include analytical, numerical, and hybrid models for the powder stream, melt pool dynamics, or part scale response, is provided in~\cite{Guan2020,Svetlizky2021}. Moreover, computational fluid dynamics models have been proposed that consider the interaction of (discrete) powder particles with the melt pool~\cite{Haley2019,Wang2021}. Most computational modeling approaches have been proposed for PBF. Existing models are based, among others, on finite element~\cite{Khairallah2014,Martin2019,Khairallah2020,Carraturo2021}, finite difference~\cite{Lee2015}, finite volume~\cite{Geiger2009,Panwisawas2017}, Lattice Boltzmann~\cite{Korner2013,Ammer2014}, or meshfree~\cite{Russell2018,Wessels2018,Wessels2019,Weirather2019,Furstenau2020,Meier2021a,Bierwisch2021a,Bierwisch2021b} discretizations, and typically consider free-surface flow with temperature-dependent surface tension, evaporation-induced recoil pressure and heat loss as well as a Gaussian laser heat source as main driving forces.

Despite the variety of models for these different AM processes, none of the aforementioned approaches explicitly considers coupled microfluid-powder dynamics along with thermo-capillary flow and phase transitions. For example, existing modeling approaches for BJT or PBF typically assume the powder particles as spatially fixed. Moreover, while some DED modeling approaches account for mobile powder particles, these are rather described as discrete point masses and not as three-dimensional continua with distributed temperature and continuous phase transition.

To help close this gap, the present work proposes a general smoothed particle hydrodynamics (SPH) modeling framework for coupled microfluid-powder dynamics in AM involving thermo-capillary two-phase flow and reversible solid-liquid phase transitions, i.e., melting and solidification. Thereby, it relies on methodologies proposed in two recent publications of the authors~\cite{Meier2021a,Fuchs2021b}. In particular, a liquid and a gas phase are considered, which are governed by temperature-dependent surface tension and wetting effects and interact with a solid phase. The latter is assumed to consist of a substrate and mobile powder particles, each modeled as an arbitrarily-shaped mobile rigid body, that are evolved in time individually. For the scenario of laser-metal interactions, the effect of rapid evaporation is incorporated through additional mechanical and thermal interface fluxes. As novel contribution of this work, a regularization of the interface forces is proposed, which allows for smooth force evolutions during phase transitions to increase the robustness of the computational scheme.

To the best of the authors' knowledge, the proposed modeling framework is the first of its kind considering coupled microfluid-powder dynamics along with the aforementioned thermo-capillary phase transition phenomena, and, thus, it is expected to become a valuable tool for detailed simulation-based studies of AM processes. To this end, the generality and robustness of the computational modeling framework is demonstrated by several application-motivated examples representing the specific AM processes binder jetting, material jetting, directed energy deposition, and powder bed fusion. Among others, it is shown how the dynamic impact of droplets in binder jetting or the evaporation-induced recoil pressure in powder bed fusion leads to powder motion, distortion of the powder packing structure, and particle ejection.

The remainder of this work is organized as follows: The governing equations for a coupled microfluid-powder dynamics problem involving thermo-capillary flow and phase transitions are outlined. Next, the computational modeling approach using SPH is presented. Finally, several numerical examples in three dimensions are considered with a focus on the specific AM processes binder jetting, material jetting, directed energy deposition, and powder bed fusion.

\section{Governing equations} \label{sec:goveq}

Within the scope of this work, two-phase flow problems of a liquid phase (with domain~$\Omega^{l}$) and a gas phase (with domain~$\Omega^{g}$) are considered that interact with a solid phase (with domain~$\Omega^{s}$). The solid phase is assumed to consist of a substrate and several arbitrarily-shaped, undeformable but mobile rigid bodies, that are evolved in time individually and allowed to get into mechanical contact with each other. Optionally, thermal conduction subject to all phases and reversible phase transitions between the solid and the liquid phase is considered. Accordingly, the overall domain splits to $\Omega = \Omega^{l} \cup \Omega^{g} \cup \Omega^{s}$ and the two-phase fluid domain is given as $\Omega^{f} = \Omega^{l} \cup \Omega^{g}$. When considering thermal conduction and phase transitions additionally the combined domain $\Omega^{h} = \Omega^{l} \cup \Omega^{s}$ is introduced. In the remainder of this work, the superscripts used to denote the domains of the phases are also used to distinguish between quantities related to these phases. Besides, a double or triple of these superscripts denote quantities related to the interface between two phases respectively the triple line between three phases. In the context of metal AM modeling, the solid phase and the gas phase correspond to the solid metal and the atmospheric gas in the build chamber of an AM device. Furthermore, depending on the considered AM process, the liquid phase either corresponds to the molten metal or the liquid binder. In the following, the governing equations for the general formulation are delineated on the basis of the authors' previous work~\cite{Fuchs2021a,Fuchs2021b,Meier2021a}.

\subsection{Fluid phases} \label{subsec:goveq_fluid}

The liquid and gas phase are governed by the instationary and anisothermal Navier-Stokes equations in the domain~$\Omega^{f}$, which consist of the continuity equation and the momentum equation in convective form
\begin{equation} \label{eq:fluid_conti}
\dv{\rho}{t} = -\rho \div{\vectorbold{u}} \qin \Omega^{f} \, ,
\end{equation}
\begin{equation} \label{eq:fluid_momentum}
\dv{\vectorbold{u}}{t} = \frac{1}{\rho} \qty( -\grad{p} + \vectorbold{f}_{\nu} + \tilde{\vectorbold{f}}^{lg}_{s} + \tilde{\vectorbold{f}}^{slg}_{w} + \tilde{\vectorbold{f}}^{lg}_{v} ) + \vectorbold{b} \qin \Omega^{f} \, .
\end{equation}
Contributions from viscous forces~$\vectorbold{f}_{\nu}$, surface tension forces~$\tilde{\vectorbold{f}}^{lg}_{s}$, wetting forces~$\tilde{\vectorbold{f}}^{slg}_{w}$, and evaporation-induced recoil pressure forces~$\tilde{\vectorbold{f}}^{lg}_{v}$, each per unit volume, as well as body forces~$\vectorbold{b}$ per unit mass, can be identified. For (incompressible) Newtonian fluids the viscous forces as defined above read $\vectorbold{f}_{\nu} = \eta \laplacian{\vectorbold{u}}$ with dynamic viscosity~$\eta$. The remaining contributions to the momentum equation are introduced below. Following a weakly compressible approach, density~$\rho$ and pressure~$p$ are linked via the equation of state
\begin{equation} \label{eq:fluid_eos}
p\qty(\rho) = c^{2} \qty(\rho - \rho_{0}) \qin \Omega^{f} \, ,
\end{equation}
with reference density~$\rho_{0}$ and artificial speed of sound~$c$. Accordingly, the reference pressure can be identified as $p_{0} = \rho_{0} c^{2}$. The Navier-Stokes equations~\eqref{eq:fluid_conti} and~\eqref{eq:fluid_momentum} are subject to initial and boundary conditions as defined in~\cite{Meier2021a}.

\subsubsection{Surface tension and wetting forces} \label{subsec:goveq_fluid_surfacetension_wetting}

Following the continuum surface force (CSF) approach~\cite{Brackbill1992}, surface tension and wetting effects are considered as volumetric forces distributed across an interfacial volume of finite width instead of additional boundary conditions at the liquid-gas interface and the solid-liquid-gas triple line. The distributed surface tension forces on the liquid-gas interface consist of the following two contributions in interface normal and tangential direction
\begin{equation} \label{eq:fluid_surfacetension}
\tilde{\vectorbold{f}}^{lg}_{s} = \alpha \kappa \vectorbold{n}^{lg} \delta^{lg} + \qty( \vectorbold{I} - \vectorbold{n}^{lg} \otimes \vectorbold{n}^{lg} ) \grad{\alpha} \delta^{lg} \, ,
\end{equation}
with the surface tension coefficient~$\alpha$, the interface curvature~$\kappa = \div \vectorbold{n}^{lg}$, and the liquid-gas interface normal~$\vectorbold{n}^{lg}$ as well as the surface delta function~$\delta^{lg}$ between liquid and gas phase. The surface delta function~$\delta^{lg}$ is employed to distribute surface tension forces across interface domains of finite thickness. It is non-zero only on these interface domains and its integral is normalized to one. A purely linear temperature-dependent surface tension coefficient, i.e., $\grad{\alpha} = \alpha'\qty(T) \grad{T}$ with $\alpha'\qty(T) = \flatfrac{\dd{\alpha\qty(T)}}{\dd{T}}$, is considered according to
\begin{equation} \label{eq:fluid_surfacetension_tempdepend}
\alpha\qty(T) = \alpha_{0} + \alpha'_{0} \qty(T - T_{\alpha_{0}}) \, ,
\end{equation}
where $\alpha_{0}$ is the surface tension coefficient at reference temperature $T_{\alpha_{0}}$ and $\alpha'_{0} < 0$. To avoid negative values of the surface tension coefficient in the regime of high temperatures, additionally the limiting condition $\alpha\qty(T) > \alpha_{min}$ is enforced. Moreover, the wetting forces acting on the solid-liquid-gas triple line are given by
\begin{equation} \label{eq:fluid_wetting}
\tilde{\vectorbold{f}}^{slg}_{w} = \alpha \qty( \cos{\theta} - \cos{\theta_{0}} ) \vectorbold{t}^{sf} \delta^{lg} \delta^{sf} \, ,
\end{equation}
with the equilibrium wetting angle~$\theta_{0}$ and the current wetting angle~$\theta$ defined via $\cos{\theta} = \vectorbold{n}^{lg} \cdot \vectorbold{n}^{sf}$, as well as the solid-fluid interface normal~$\vectorbold{n}^{sf}$ and tangent~$\vectorbold{t}^{sf}$, and the surface delta function~$\delta^{sf}$ between solid and fluid phase.

\subsubsection{Evaporation-induced recoil pressure forces} \label{subsec:goveq_fluid_recoil}

In PBF and DED processes, the high peak temperatures at typical process conditions give rise to considerable evaporation effects. Herein, a phenomenological model for the evaporation-induced recoil pressure forces based on~\cite{Anisimov1995} is employed
\begin{equation} \label{eq:fluid_recoil}
\tilde{\vectorbold{f}}^{lg}_{v} = -p_{v}\qty(T) \vectorbold{n}^{lg} \delta^{lg}
\qq{with}
p_{v}\qty(T) = C_{P} \exp \qty[ - C_{T} \qty( \frac{1}{T} - \frac{1}{T_{v}} ) ] \, ,
\end{equation}
where the constants $C_{P} = 0.54 p_{a}$ as well as $C_{T} = \flatfrac{\bar{h}_{v}}{R}$ contain the atmospheric pressure~$p_{a}$, the molar latent heat of evaporation~$\bar{h}_{v}$, and the molar gas constant~$R$. With typical parameter values for liquid metals~\cite{Khairallah2016, Weirather2019} at an ambient pressure of $p_{a} = \SI{e5}{\newton\per\meter\squared}$, these constants take on values of $C_{P} = \SI{5.4e4}{\newton\per\meter\squared}$ and $C_{T} \approx \SI{5.0e4}{\kelvin}$. Moreover, $T_{v}$ is the boiling temperature of the liquid phase.

\subsection{Solid phase} \label{subsec:goveq_solid}

The solid phase is assumed to consist of a substrate and several arbitrarily-shaped, undeformable but mobile rigid bodies that are each represented by a sub-domain of~$\Omega^{s}$ and embedded in the fluid domain~$\Omega^{f}$. The motion of an individual rigid body is described by the balance of linear and angular momentum, among others, considering coupling contributions from interaction with the fluid phase and from mechanical contact with the substrate and neighboring rigid bodies. This allows to evolve each rigid body in time individually. For the sake of brevity, the equations of motion are not delineated herein, referring to~\cite{Fuchs2021b} instead.

\subsection{Thermal conduction} \label{subsec:goveq_thermal}

The fluid and the solid phases are subject to thermal conduction that is governed by the following energy equation as
\begin{equation} \label{eq:thermal_energy}
c_{p} \dv{T}{t} = \frac{1}{\rho} \qty( -\div{\vectorbold{q}} + \tilde{s}^{hg}_{l} + \tilde{s}^{lg}_{v} ) \qin \Omega \, ,
\end{equation}
with specific heat capacity~$c_{p}$. The heat flux is defined as $\vectorbold{q} = -k \grad{T}$ according to Fourier's law with thermal conductivity~$k$. The energy equation~\eqref{eq:thermal_energy} is subject to initial and boundary conditions as defined in~\cite{Meier2021a}.

\begin{rmk}
Note that thermal convection is hidden in the total time derivative~$\dv*{T}{t}$ of the energy equation~\eqref{eq:thermal_energy} written Lagrangian description.
\end{rmk}

Besides, heat fluxes stemming from the laser beam heat source~$\tilde{s}^{hg}_{l}$ and from evaporation-induced heat losses~$\tilde{s}^{lg}_{v}$, each per unit volume, are considered. The former is given by
\begin{equation} \label{eq:fluid_heatsource}
\tilde{s}^{hg}_{l} = \zeta_{l} \expval{-\vectorbold{n}^{hg} \cdot \vectorbold{e}_{l}} s^{hg}_{l}\qty(\vectorbold{x}) \delta^{hg}
\qq{with}
s^{hg}_{l}\qty(\vectorbold{x}) = s^{hg}_{l0} \exp\qty[-2 \qty( \frac{\norm{\vectorbold{x}-\vectorbold{x}_{0}}}{r_w} )^{2} ] \, ,
\end{equation}
where $\zeta_{l}$ is the laser energy absorptivity and $\vectorbold{e}_{l}$ is the unit vector representing the laser beam direction. The Macauley bracket $\expval{\cdot}$ returns the value of its argument if the argument is positive and zero otherwise. The irradiance $s^{hg}_{l}\qty(\vectorbold{x})$ describes the incident laser power per unit area at position $\vectorbold{x}$ as a function of the laser beam center position $\vectorbold{x}_{0}$ and has the form of a Gaussian distribution with the peak value $s^{hg}_{l0}$ and the standard deviation $\sigma = \flatfrac{r_{w}}{2}$. The quantity $d_{w} = 2 r_{w}$ is a frequently used measure for the effective laser beam diameter. With a given total laser power~$P_{l}$ the peak value follows from normalization as~$s^{hg}_{l0} = \flatfrac{2 P_{l}}{\pi r_{w}^{2}}$. Eventually, following the same phenomenological model as for the recoil pressure forces~\eqref{eq:fluid_recoil}, the evaporation-induced heat loss reads
\begin{equation} \label{eq:fluid_evaporation}
\tilde{s}^{lg}_{v} = s^{lg}_{v} \delta^{lg}
\qq{with}
s^{lg}_{v} = - \dot{m}^{lg}_{v} \qty(h_{v} + h\qty(T)) \, ,
\quad
\dot{m}^{lg}_{v} = 0.82 c_{s} p_{v}\qty(T) \sqrt{\frac{C_{M}}{T}} \, ,
\quad
h\qty(T) = \int\limits_{T_{h,0}}^T c_{p} \dd{\bar{T}} \, ,
\end{equation}
where the enthalpy rate per unit area~$s^{lg}_{v}$ results from the vapor mass flow per unit area~$\dot{{m}}^{lg}_{v}$ and the sum of the specific enthalpy~$h(T)$ and the latent heat of evaporation~$h_{v}$, both per unit mass. Moreover, $T_{h,0}$ is a reference temperature of the specific enthalpy and the constant $C_{M} = \flatfrac{M}{(2\pi R)}$ contains the molar mass~$M$ and the molar gas constant~$R$. Finally, $p_{v}\qty(T)$ is the recoil pressure defined in~\eqref{eq:fluid_recoil} and $c_{s}$ the so-called sticking constant which takes on a value close to one for metals~\cite{Khairallah2016, Weirather2019}.

\section{Spatial discretization via smoothed particle hydrodynamics} \label{sec:nummeth_sph}

For the spatial discretization of the governing equations the method of SPH is used. The SPH formulation applied herein combines two of the authors' approaches recently proposed in the literature: First, a weakly compressible SPH formulation modeling multiphase fluid flow for thermo-capillary phase transition problems~\cite{Meier2021a}. Second, a fully resolved SPH formulation for fluid-solid and contact interaction problems including thermo-mechanical coupling and reversible phase transitions~\cite{Fuchs2021b}. Accordingly, this section gives a brief overview of the combined SPH formulation based on~\cite{Fuchs2021a,Fuchs2021b,Meier2021a} and proposes additional methodological novelties needed to model a host of complex multiphysics problems in the field of metal AM in an accurate and robust manner. The proposed modeling framework is implemented in the in-house parallel multiphysics research code BACI (Bavarian Advanced Computational Initiative)~\cite{Baci}.

\begin{rmk}
The domain $\Omega$ is initially discretized by particles located on a regular grid with spacing~$\Delta{}x$. The mass of a particle is assigned using its reference density~$\rho_{0}$ and its effective volume~$\qty(\Delta{}x)^{d}$ (given in $d$-dimensional space). To introduce a short notation, a quantity~$f$ evaluated for particle~$i$ at position~$\vectorbold{r}_{i}$ is written as~$f_{i} = f\qty(\vectorbold{r}_{i})$. Besides, $W_{ij} = W\qty(r_{ij}, h)$ denotes the smoothing kernel~$W$ evaluated for particle~$i$ at position~$\vectorbold{r}_{i}$ with neighboring particle~$j$ at position~$\vectorbold{r}_{j}$, where $r_{ij} = \norm{\vectorbold{r}_{i} - \vectorbold{r}_{j}}$ is the absolute distance between particles~$i$ and~$j$, and $h$ is the smoothing length. Similarly, the derivative of the smoothing kernel~$W$ with respect to the absolute distance~$r_{ij}$ is denoted by $\pdv*{W}{r_{ij}} = \pdv*{W\qty(r_{ij}, h)}{r_{ij}}$. The initial particle spacing~$\Delta{}x$ is set equal to the smoothing length~$h$. Besides, a quintic spline smoothing kernel~$W\qty(r, h)$~\cite{Morris1997} with smoothing length~$h$ and support radius~$r_{c} = 3 h$ is applied.
\end{rmk}

\subsection{Discretization of phase interfaces} \label{subsec:nummeth_sph_colorfield}

The representation of different phase interfaces is crucial for the evaluation of mechanical interface forces or thermal interface heat fluxes. To this end, a density-weighted color field function and its gradient are defined based on~\cite{Adami2010,Breinlinger2013} as
\begin{equation} \label{eq:sph_colorfield}
c_{i} = \frac{1}{V_{i}} \sum_{j} \chi_{j} \qty(V_{i}^{2}+V_{j}^{2}) \frac{\rho_{i}}{\rho_{i}+\rho_{j}} W_{ij}
\qand
\grad{c}_{i} = \frac{1}{V_{i}} \sum_{j} \chi_{j} \qty(V_{i}^{2}+V_{j}^{2}) \frac{\rho_{i}}{\rho_{i}+\rho_{j}} \pdv{W}{r_{ij}} \vectorbold{e}_{ij} \, ,
\end{equation}
evaluated for particles $i$ and $j$ belonging to different phases. In addition to~\cite{Adami2010,Breinlinger2013}, an optional scaling factor~$\chi_{j}$ is introduced, that is by default set to one unless stated otherwise. Based on the definition of the color field gradient~\eqref{eq:sph_colorfield} the surface delta function and the interface normal of particle~$i$ read
\begin{equation} \label{eq:sph_normalanddelta}
\delta_{i} = \norm{\grad{c}_{i}}
\qand
\vectorbold{n}_{i} =
\begin{cases}
\flatfrac{\grad{c}_{i}}{ \norm{\grad{c}_{i}} } & \qq*{if} \norm{\grad{c}_{i}} > \epsilon \, , \\
\vectorbold{0} & \qq*{otherwise.}
\end{cases}
\end{equation}
This procedure leads to an outward-pointing interface normal~$\vectorbold{n}_{i}$ with respect to the phase of particle~$i$. The tolerance $\epsilon \ll 1$ is applied to avoid erroneous interface normals for particles far away from the interface. Note that the metrics~\eqref{eq:sph_colorfield}-\eqref{eq:sph_normalanddelta} are, by definition, exclusively used to represent the interface between two phases and not the triple line between three phases.

\begin{rmk}
In case of high density ratios between two phases the definition of $\delta_{i}$ according to~\eqref{eq:sph_colorfield} and~\eqref{eq:sph_normalanddelta} ensures that the majority of a flux contribution, i.e., of mechanical interface forces or thermal heat fluxes, distributed over the interface via $\delta_{i}$ acts on the phase associated with the higher density.
\end{rmk}

\subsection{Phase transitions and treatment of resultant discontinuities} \label{subsec:nummeth_sph_discontinuities}

Reflecting the Lagrangian nature of SPH, each particle carries its phase information. Accordingly, the discretized energy equation~\eqref{eq:sph_energy} is evaluated for each particle with phase-specific parameters. Furthermore, particles undergo phase transitions solid~$\leftrightarrow$~liquid when exceeding or falling below the melt temperature~$T_{m}$. Due to this procedure, discontinuities in time occur that originate from abruptly changing contributions to the discretized momentum equation~\eqref{eq:sph_momentum} after particles change phase from solid to liquid or vice versa. The primary driving forces of thermo-capillary phase transition problems are typically surface tension and wetting forces, while viscous and gravity forces can be considered as secondary effects~\cite{Meier2017b}. The discontinuities as described above shall be avoided by employing a regularization procedure focusing on relevant terms of these primary driving forces. To this end, a linear transition function~\eqref{eq:sph_transition} is introduced to scale the respective contributions of particles close to the melt temperature~$T_{m}$ with the goal to smoothen out these discontinuities.

\begin{rmk}
The linear transition function $f\qty[x, x_{1}, x_{2}]$ and the complementary linear transition function $\bar{f}\qty[x, x_{1}, x_{2}] = 1-f\qty[x, x_{1}, x_{2}]$ with arbitrary arguments $x$, $x_{1} < x_{2}$ are defined as
\begin{equation} \label{eq:sph_transition}
f\qty[x, x_{1}, x_{2}] =
\begin{cases}
1 & \qq*{if} x > x_{2} \, , \\
\frac{x-x_{1}}{x_{2}-x_{1}} & \qq*{if} x_{2} \geq x \geq x_{1} \, , \\
0 & \qq*{if} x < x_{1} \, ,
\end{cases}
\qand
\bar{f}\qty[x, x_{1}, x_{2}] =
\begin{cases}
0 & \qq*{if} x > x_{2} \, , \\
\frac{x_{2}-x}{x_{2}-x_{1}} & \qq*{if} x_{2} \geq x \geq x_{1} \, , \\
1 & \qq*{if} x < x_{1} \, .
\end{cases}
\end{equation}
\end{rmk}

\begin{rmk}
The vapor phase is not modeled explicitly herein such that phase transitions liquid~$\leftrightarrow$~vapor are considered implicitly in terms of evaporation-induced recoil pressure forces~\eqref{eq:fluid_recoil} and heat losses~\eqref{eq:fluid_evaporation}. Note that the latent heat of melting could be considered in a straightforward manner as well by employing, e.g., an apparent capacity scheme relying on an increased heat capacity $c_{p}$ within a finite temperature interval~\cite{Proell2020}.
\end{rmk}

\subsection{Modeling microfluidic flow via weakly compressible SPH} \label{subsec:nummeth_sph_fluid}

The density of a particle~$i$ is determined via summation $\rho_{i} = m_{i} \sum_{j} W_{ij}$ of the respective smoothing kernel contributions of all neighboring particles~$j$ resulting in exact conservation of mass. Hence, the actual volume of a particle~$i$ is computed as~$V_{i} = \flatfrac{m_{i}}{\rho_{i}}$ and the pressure~$p_{i}$ follows directly from evaluation of the discrete version of the equation of state~\eqref{eq:fluid_eos} for particle~$i$. The discretized momentum equation to evaluate the total acceleration~$\vectorbold{a}_{i} = \dv*{\vectorbold{u}_{i}}{t}$ of a particle~$i$ can be formulated as
\begin{equation} \label{eq:sph_momentum}
\vectorbold{a}_{i} = \frac{1}{m_{i}} \qty( \vectorbold{F}_{p,i} + \vectorbold{F}_{\nu,i} + \vectorbold{F}_{s,i} + \vectorbold{F}_{v,i} + \vectorbold{F}_{d,i} + \vectorbold{F}_{b,i}) + \vectorbold{b}_{i} \, .
\end{equation}
The pressure forces~$\vectorbold{F}_{p,i}$, viscous forces~$\vectorbold{F}_{\nu,i}$, surface tension forces~$\vectorbold{F}_{s,i}$ as well as evaporation-induced recoil pressure forces~$\vectorbold{F}_{v,i}$ acting on particle~$i$ result from summation of all interaction contributions with neighboring particles~$j$. Optionally, additional viscous dissipation forces~$\vectorbold{F}_{d,i}$ are applied at the liquid-gas and solid-fluid interface, and, barrier forces~$\vectorbold{F}_{b,i}$ are acting in rare scenarios of impermissibly close particles only to retain reasonable particle distributions. No-slip boundary conditions are modeled using a boundary particle formulation proposed in~\cite{Adami2012}. In the following, the force contributions in the momentum equation above including regularization procedures to avoid discontinuities stemming from phase transitions solid~$\leftrightarrow$~liquid are discussed.

\subsubsection{Pressure and viscous forces} \label{subsec:nummeth_sph_pressure_viscous_force}

As in the authors' previous works~\cite{Fuchs2021a,Fuchs2021b,Meier2021a}, the pressure and viscous forces are discretized following a formulation proposed in~\cite{Adami2012, Adami2013} as
\begin{equation} \label{eq:sph_momentum_pressure_and_viscous}
\vectorbold{F}_{p,i} + \vectorbold{F}_{\nu,i} = \sum_{j} \qty(V_{i}^{2}+V_{j}^{2}) \qty( - \frac{\rho_{j}p_{i}+\rho_{i}p_{j}}{\rho_{i} + \rho_{j}} \pdv{W}{r_{ij}} \vectorbold{e}_{ij} + \frac{2\eta_{i}\eta_{j}}{\eta_{i}+\eta_{j}} \frac{\vectorbold{u}_{ij}}{r_{ij}} \pdv{W}{r_{ij}} ) \, ,
\end{equation}
with unit vector $\vectorbold{e}_{ij} = \flatfrac{\vectorbold{r}_{i} - \vectorbold{r}_{j}}{r_{ij}}$ and relative velocity $\vectorbold{u}_{ij} = \vectorbold{u}_{i}-\vectorbold{u}_{j}$. In addition, the transport-velocity formulation~\cite{Adami2013}, which utilizes a constant background pressure $p_{b}$ to suppress the problem of tensile instability is employed herein.

\subsubsection{Surface tension forces} \label{subsec:nummeth_sph_surface_tension}

The surface tension forces consist of an interface normal curvature-proportional contribution and an interface tangential Marangoni contribution due to surface tension gradients. Accordingly, the surface tension forces are given as
\begin{equation} \label{eq:sph_surfacetension}
\vectorbold{F}_{s,i} = f\qty[T_{i}, T_{m}, T_{m} + \Delta{}T_{s}] \qty( - V_{i} \alpha_{i} \kappa_{i} \grad{c}_{i}^{lg} + V_{i} \alpha_{i}' \delta^{lg}_{i} \grad_{t} {T}_{i} ) \, ,
\end{equation}
with the curvature
\begin{equation} \label{eq:sph_curvature}
\kappa_{i} = - \frac{ \sum_{j} f\qty[T_{j}, T_{m}, T_{m} + \Delta{}T_{s}] \cdot V_{j} \qty(\vectorbold{n}^{lg}_{i}-\vectorbold{n}^{lg}_{j}) \pdv{W}{r_{ij}} \vectorbold{e}_{ij} }{ \sum_{j} f\qty[T_{j}, T_{m}, T_{m} + \Delta{}T_{s}] \cdot V_{j} W_{ij} }
\end{equation}
based on a formulation proposed in~\cite{Morris2000}. Here, $\grad_{t} {T}_{i}$ represents the projection of the temperature gradient into the interface tangential plane following
\begin{equation} \label{eq:sph_tempgrad}
\grad_{t} {T}_{i} = \qty( \vectorbold{I} - \vectorbold{n}^{lg}_{i} \otimes \vectorbold{n}^{lg}_{i} ) \grad{T}_{i}
\qq{with}
\grad{T}_{i} = \sum_{j} V_{j} (T_{j} - T_{i}) \pdv{W}{r_{ij}} \vectorbold{e}_{ij} \, .
\end{equation}
The surface tension forces~\eqref{eq:sph_surfacetension} are computed for particles $i$ and $j$ of the liquid and gas phase. To avoid discontinuities stemming from phase transitions solid~$\leftrightarrow$~liquid, the surface tensions forces~\eqref{eq:sph_surfacetension} are regularized within a small temperature interval~$\Delta{}T_{s}$ above the melt temperature~$T_{m}$ utilizing the linear transition function~\eqref{eq:sph_transition}. Likewise, the contributions from neighboring particles~$j$ to the curvature~\eqref{eq:sph_curvature} are regularized. Note that in the evaluation of the temperature gradient~\eqref{eq:sph_tempgrad} particles of all phases, i.e., solid, liquid, and gas, are considered. For this reason, the temperature gradient will not suffer from these discontinuities.

\subsubsection{Wetting phenomena} \label{subsec:nummeth_sph_wetting}

Following a strategy proposed in~\cite{Breinlinger2013}, the wetting forces~\eqref{eq:fluid_wetting} are not discretized and evaluated directly. Instead, the desired equilibrium wetting angle~$\theta_{0}$ is enforced prescribing the interface normal between the liquid and gas phase in the triple line region. To begin with, the standard interface normal between liquid and gas phase is evaluated following~\eqref{eq:sph_colorfield} and~\eqref{eq:sph_normalanddelta} and denoted as $\bar{\vectorbold{n}}^{lg}_{i}$ in the following. In a next step, a distance measure and an interface normal for a particle~$i$ of the fluid phase with respect to particles $j$ of the solid phase are defined. Thereto, the color field function~$c_{i}^{sf}$ as well as its gradient~$\grad{c}_{i}^{sf}$ at the solid-fluid interface are defined based on~\eqref{eq:sph_colorfield} with the scaling factor~$\chi_{j} = \bar{f}\qty[T_{j}, T_{m} - \Delta{}T_{s}, T_{m}]$. The additional factor ensures that for a particle~$i$ the color field function and also its gradient do not change abruptly when particles~$j$ are subject to phase transitions solid~$\leftrightarrow$~liquid. The corresponding interface normal~$\vectorbold{n}_{i}^{sf}$ is defined based on~\eqref{eq:sph_normalanddelta}. The interface normal to be prescribed in the triple line region between the liquid and gas phase is determined on the basis of the equilibrium wetting angle $\theta_{0}$ according to
\begin{equation} \label{eq:sph_wetting1}
\hat{\vectorbold{n}}^{lg}_{i} = \vectorbold{t}^{sf}_{i} \sin{\theta_{0}} - \vectorbold{n}_{i}^{sf} \cos{\theta_{0}} \, ,
\end{equation}
where the interface tangent $\vectorbold{t}^{sf}_{i}$ is given by
\begin{equation} \label{eq:sph_wetting2}
\vectorbold{t}^{sf}_{i} =
\begin{cases}
\flatfrac{\tilde{\vectorbold{t}}^{sf}_{i}}{\norm{\tilde{\vectorbold{t}}^{sf}_{i}}} & \qq*{if} \norm{\tilde{\vectorbold{t}}^{sf}_{i}} > \epsilon \, , \\
\vectorbold{0} & \qq*{otherwise,}
\end{cases}
\qq{with}
\tilde{\vectorbold{t}}^{sf}_{i} = \bar{\vectorbold{n}}^{lg}_{i} - \qty(\bar{\vectorbold{n}}^{lg}_{i} \cdot \vectorbold{n}^{sf}_{i}) \vectorbold{n}^{sf}_{i}.
\end{equation}
The denominator above can only become zero when the interface normals $\bar{\vectorbold{n}}^{lg}_{i}$ and $\vectorbold{n}^{sf}_{i}$ are parallel or anti-parallel, which only 
happens for very rare configurations, e.g., perfect wetting, or a liquid bubble close to a solid wall with a very thin gas film in between. To have a smooth transition of the liquid-gas interface normal~$\vectorbold{n}^{lg}_{i}$ from the triple line region with prescribed interface normal~$\hat{\vectorbold{n}}^{lg}_{i}$ to the interior domain with standard interface normal~$\bar{\vectorbold{n}}^{lg}_{i}$, the following correction scheme in analogy to~\cite{Breinlinger2013} is employed:
\begin{equation} \label{sph:fluid_wetting3}
\vectorbold{n}^{lg}_{i} = 
\begin{cases}
\flatfrac{\tilde{\vectorbold{n}}^{lg}_{i} }{\norm{\tilde{\vectorbold{n}}^{lg}_{i}}} & \qq*{if} \norm{\tilde{\vectorbold{n}}^{lg}_{i}} > \epsilon \, , \\
\vectorbold{0} & \qq*{otherwise,}
\end{cases}
\qq{with}
\tilde{\vectorbold{n}}^{lg}_{i} = f\qty[{c}_{i}^{sf}, c_{1}, c_{2}] \cdot \hat{\vectorbold{n}}^{lg}_{i} + \bar{f}\qty[{c}_{i}^{sf}, c_{1}, c_{2}] \cdot \bar{\vectorbold{n}}^{lg}_{i}
\end{equation}
As long as the norms of $\hat{\vectorbold{n}}^{lg}_{i}$ and $\bar{\vectorbold{n}}^{lg}_{i}$ are equal to unity, the denominator can only be zero if the two vectors are anti-parallel, which is very unlikely for configurations sufficiently close to the equilibrium wetting angle~$\theta_{0}$. Note that the correction parameters $c_{1} = 0.0$ and $c_{2} = 0.2$ were found to be reasonable.

\subsubsection{Recoil pressure forces} \label{subsec:nummeth_sph_recoil}

The discrete version of the evaporation-induced recoil pressure forces~\eqref{eq:fluid_recoil} based on a phenomenological model~\cite{Anisimov1995} is given by
\begin{equation} \label{eq:sph_recoil}
\vectorbold{F}_{v,i} = - V_{i} p_{v,i} \vectorbold{n}_{i}^{lg} \delta_{i}^{lg},
\end{equation}
where $p_{v,i}$ is the recoil pressure according to~\eqref{eq:fluid_recoil} evaluated for particle~$i$.

\subsubsection{Viscous dissipation forces} \label{subsec:nummeth_sph_dissipation}

As proposed in~\cite{Meier2021a}, viscous dissipation forces are employed selectively at the liquid-gas and solid-fluid interface to avoid oscillations originating from the primary force contributions at the liquid-gas interface and from the phase transitions solid~$\leftrightarrow$~liquid. The discrete version of the viscous dissipation forces is based on a stabilization term denoted as artificial viscosity~\cite{Monaghan1983} and is given as
\begin{equation} \label{eq:sph_dissipation}
\vectorbold{F}_{d,i} = - m_{i} \alpha_{i} \sum_{j} m_{j} \frac{ h \bar{c}_{ij}}{\bar{\rho}_{ij}} \frac{\vectorbold{u}_{ij} \cdot \vectorbold{r}_{ij}}{( r_{ij}^{2} + \epsilon h^{2} )} \pdv{W}{r_{ij}} \, ,
\end{equation}
with the inter-particle averaged speed of sound $\bar{c}_{ij} = \flatfrac{(c_{i}+c_{j})}{2}$ and density $\bar{\rho}_{ij} = \flatfrac{(\rho_{i}+\rho_{j})}{2}$. Here, the constant $\epsilon \ll 1$ is applied to ensure a non-zero denominator. The viscosity factor is split into two contributions $\alpha_{i} = \alpha^{lg}_{i}+\alpha^{sf}_{i}$. The first one is given as
\begin{equation} \label{eq:sph_dissipation_lg}
\alpha^{lg}_{i} = \alpha^{lg}_{0} h \delta^{lg}_{i}
\end{equation}
and is acting only on the liquid-gas interface. It has been demonstrated in~\cite{Meier2021a}, that this contribution (if $\alpha^{lg}_{0}$ is chosen in a reasonable range) effectively reduces spurious interface flows known to occur in CSF formulations~\cite{Brackbill1992,Breinlinger2013} without introducing additional dissipation of physically relevant flow characteristics in the interior fluid domain. The second contribution of the viscosity factor is defined as
\begin{equation} \label{eq:sph_dissipation_sf}
\alpha^{sf}_{i} = \alpha^{sf}_{0} \bar{f}\qty[T_{i}, T_{m}, T_{m} + \Delta{}T_{d}] \, .
\end{equation}
Accordingly, this contribution only acts on particles of the fluid phase with temperatures close to the melt temperature~$T_{m}$ and thus effectively damps potential instabilities arising from phase transitions solid~$\leftrightarrow$~liquid. Both contributions to the viscosity factor can also be interpreted from a physical point of view as a non-conservative surface tension formulation with interface viscosity~\cite{Gounley2016} and a physical model for the gradual phase transition of alloys between liquidus and solidus temperature.

\subsubsection{Barrier forces} \label{subsec:nummeth_sph_barrier}

In order to retain reasonable particle distributions and to circumvent instabilities, especially in the triple line region during phase transitions, additional elastic and viscous barrier forces are introduced. These barrier forces only act in rare scenarios where particles are impermissibly close to each other and thus have no significant effect on the physical flow characteristics of the problem. The barrier forces are defined according to
\begin{equation} \label{sph:barrier}
\vectorbold{F}_{b,i} = \sum_{j} \qty( k_{b} g_{ij} + d_{b} \abs{g_{ij}} \dot{g}_{ij} ) \vectorbold{e}_{ij}
\qq{with}
g_{ij} =
\begin{cases}
r_{b}-r_{ij} & \qq*{if} r_{ij} < r_{b} \, , \\
0 & \qq*{otherwise,}
\end{cases}
\quad
\dot{g}_{ij} =
\begin{cases}
- \vectorbold{e}_{ij} \cdot \vectorbold{u}_{ij} & \qq*{if} r_{ij} < r_{b} \, , \\
0 & \qq*{otherwise,}
\end{cases}
\end{equation}
with stiffness constant $k_{b}$, damping constant $d_{b}$, and interaction distance $r_{b} < h$.

\subsection{Modeling the motion of rigid bodies discretized by particles} \label{subsec:nummeth_sph_solid}

As proposed in the authors' previous work~\cite{Fuchs2021b}, the rigid bodies making up the solid phase are fully resolved, that is spatially discretized as clusters of particles. This approach offers several advantages: First, advanced boundary particle methods, e.g., based on the extrapolation of fluid quantities to the solid phase~\cite{Morris1997,Basa2009,Adami2012}, to accurately model momentum exchange at the fluid-solid interface, can be utilized. Second, thermal conduction in the solid and the fluid phase can be modeled, e.g., based on~\cite{Cleary1999}, without the need for sophisticated coupling formulations at the fluid-solid interface. Finally, phase transitions solid~$\leftrightarrow$~liquid can be evaluated in a straightforward manner for each involved particle separately. For more details on the modeling approach refer to~\cite{Fuchs2021b}.

\subsection{Modeling thermal conduction} \label{subsec:nummeth_sph_energy}

The conductive term in the energy equation~\eqref{eq:thermal_energy} is discretized using a formulation from~\cite{Cleary1999}, that is especially suitable for problems involving a discontinuity of the thermal conductivity $k$ across phase interfaces. Accordingly, the discretized form of the energy equation reads
\begin{equation} \label{eq:sph_energy}
c_{p,i} \dv{T_{i}}{t} = \frac{1}{\rho_{i}} \qty( \sum_{j} V_{j} \frac{4k_{i}k_{j}}{k_{i}+k_{j}} \frac{T_{j}-T_{i}}{r_{ij}} \pdv{W}{r_{ij}} + \tilde{s}^{hg}_{l,i} + \tilde{s}^{lg}_{v,i}) \, .
\end{equation}
The discrete versions of the laser beam source term $\tilde{s}^{hg}_{l,i}$ and the evaporation-induced heat loss term $\tilde{s}^{lg}_{v,i}$ result directly from 
evaluating~\eqref{eq:fluid_heatsource} and~\eqref{eq:fluid_evaporation} for the discrete particle~$i$.

\subsection{Time integration scheme} \label{subsec:nummeth_sph_timint}

The discretized governing equations are integrated in time applying explicit time integration schemes. For the sake of brevity, these schemes are not further delineated herein referring to~\cite{Fuchs2021a,Fuchs2021b,Meier2021a} instead. Several restrictions on the time step size~$\Delta{}t$ are required to maintain stability of the applied time integration schemes, which are the Courant-Friedrichs-Lewy (CFL) condition, the viscous condition, the body force condition, the surface tension condition, the contact condition, and the conductivity-condition~\cite{Morris1997,Adami2013,Adami2010,OSullivan2004,Cleary1998}
\begin{equation} \label{eq:sph_timestepcond}
\Delta{}t \leq \min\qty{
0.25 \frac{h}{c+\norm{\vectorbold{u}_{max}}}, \quad
0.125\frac{h^{2}}{\nu}, \quad
0.25 \sqrt{\frac{h}{\norm{\vectorbold{b}_{max}}}}, \quad
0.25 \sqrt{\frac{\rho h^3}{2 \pi \alpha}}, \quad
0.22 \sqrt{\frac{m}{k}},
\quad
0.1 \frac{\rho c_{p} h^{2}}{k}
} \, ,
\end{equation}
with maximum fluid velocity~$\vectorbold{u}_{max}$ and maximum body force~$\vectorbold{b}_{max}$. The contact condition with the mass~$m$ of involved particles is relevant for barrier forces with stiffness constant~$k_{b}$ as well as for solid contact forces with stiffness constant~$k_{c}$~\cite{Fuchs2021b}. In both cases also the damping constants~$d_{b}$ and~$d_{c}$ should be limited (here based on a trial and error approach) to avoid time integration instabilities.

\section{Numerical examples} \label{sec:results}

In the following, the versatility and robustness of the proposed computational modeling framework is demonstrated examining several numerical examples in three dimension. The focus is set on showcasing the models general capability to capture the physical phenomena that are characteristic for AM processes such as BJT, MJT, DED, and PBF. Accordingly, representative material parameters are selected, while, for the sake of simplicity, detailed process conditions are not entirely met but replicated in an idealized way.

\subsection{Representative configuration for the numerical simulation of additive manufacturing processes} \label{sec:results_config}

This section gives a representative configuration for the numerical examples in the context of the considered AM processes, that is, unless explicitly stated otherwise, applied in the subsequent examples. As in the authors' previous work~\cite{Meier2021a}, representative material parameters close to stainless steel at the melt temperature~$T_{m} = \SI{1700}{\kelvin}$ are selected based on~\cite{Khairallah2014,Khairallah2016}. To provide an overview, the material parameters for molten and solid metal applied herein are given in Tables~\ref{tab:material_params_molten_metal} and~\ref{tab:material_params_solid_metal}, while the material parameters for atmospheric gas are given in Table~\ref{tab:material_params_atmoshperic_gas}. To keep the computational effort at a feasible level considering the time step restrictions~\eqref{eq:sph_timestepcond}, the density and viscosity of atmospheric gas are selected based on the ratios $\flatfrac{\rho_{0}^{l}}{\rho_{0}^{g}} = 100$ and $\flatfrac{\eta^{l}}{\eta^{g} = 10}$ with respect to the material parameters of molten metal. In the case of BJT examples, the material parameters as given in Table~\ref{tab:material_params_liquid_binder} are applied for liquid binder. The initial powder bed as used for the BJT and PBF examples is obtained in a pre-processing step based on a cohesive powder model~\cite{Meier2019a,Meier2019b} using the discrete element method (DEM) and a log-normal type size distribution for the powder particles with diameters between $\SI{16}{\micro\meter}$ and $\SI{32}{\micro\meter}$. All phases are initially at rest with the temperature~$\SI{500}{\kelvin}$. On all boundaries no-slip conditions are applied and the temperature~$\SI{500}{\kelvin}$ is prescribed. A gravitational acceleration of magnitude~$\SI{9.81}{\meter\per\second\squared}$ is acting in downward direction set as body force per unit mass of all involved phases. A standard discretization strategy with initial particle spacing~$\Delta{}x = \SI[parse-numbers=false]{1.\overline{6}}{\micro\meter}$ and time step size~$\Delta{}t = \SI{1.0e-6}{\milli\second}$ is applied as given in Table~\ref{tab:discretization_params}. For all fluid phases, the reference pressure of the weakly compressible model is set to $p_{0} = \SI{1.0e7}{\pascal}$ and the background pressure of the transport velocity formulation~\cite{Adami2013} to $p_{b} = 5 p_{0}$. Barrier forces are evaluated with interaction distance $r_{b} = \SI[parse-numbers=false]{0.8\overline{3}}{\micro\meter}$ and the stiffness and damping constants $k_{b} = \SI{1.0}{\kilogram\per\second\squared}$ and $d_{b} = \SI{1.0e-4}{\kilogram\per\second}$. The stiffness and damping constants applied for contact evaluation of rigid bodies are set to $k_{c} = \SI{1.0}{\kilogram\per\second\squared}$ and $d_{c} = \SI{1.0e-4}{\kilogram\per\second}$. The domain boundaries are modeled as rigid walls using a boundary particle formulation following~\cite{Adami2012}. In a post-processing step the obtained particle-based results are visualized applying an SPH approximation as in the authors' previous work~\cite{Fuchs2021a,Fuchs2021b,Meier2021a}.

\subsection{Binder jetting} \label{subsec:results_bjt}

This example aims to demonstrate the general applicability of the proposed modeling framework in the context of BJT processes. For that purpose, coupled microfluid-powder dynamics are considered while thermal effects are irrelevant. Consider a domain with dimension $\SI{320}{\micro\meter} \times \SI{320}{\micro\meter} \times \SI{200}{\micro\meter}$ (discretized by approximately \num{4.5e6} SPH particles). The lower part of the domain is occupied by a substrate of thickness $\SI{20}{\micro\meter}$. A powder bed consisting of a total of 500 individual powder particles with diameters between $\SI{16}{\micro\meter}$ and $\SI{32}{\micro\meter}$ is resting on the substrate. A spherical binder droplet with diameter $\SI{80}{\micro\meter}$ is placed centrally above the powder bed. The center of the droplet is a distance of $\SI{100}{\micro\meter}$ above the substrate. The remainder of the domain is filled with atmospheric gas and surrounded by rigid walls. The binder droplet is initialized with a velocity pointing in downward direction such that it is approaching the powder bed and the substrate. In the following, three variants with different values for this velocity are considered, namely, $\SI{1}{\meter\per\second}$, $\SI{10}{\meter\per\second}$ and $\SI{50}{\meter\per\second}$. Note, that the impact velocity in the real process depends on the nozzle exit velocity of the binder droplet and the falling height.

To begin with, a time series of the results until $t = \SI{0.1}{\milli\second}$ obtained with initial velocity $\SI{1}{\meter\per\second}$ of the liquid binder droplet are given in Figures~\ref{fig:bjt_vel01_half} and~\ref{fig:bjt_vel01_full} (see Supplementary Videos 1 and 2). The liquid binder droplet gradually penetrates into the powder bed. Slight powder dynamics can be observed as metal powder particles are gently pushed outwards by the liquid binder. The liquid binder remains in a compact shape enclosing solely few metal powder particles in the center. At $t = \SI{0.1}{\milli\second}$ a downward movement of the binder towards the substrate is still visible.

\begin{figure}[htbp]
\centering
\subfigure [$t = \SI{0.000}{\milli\second}$]
{
\includegraphics[width=0.23\textwidth]{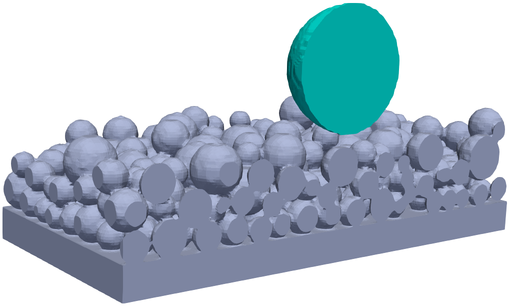}
}
\subfigure [$t = \SI{0.050}{\milli\second}$]
{
\includegraphics[width=0.23\textwidth]{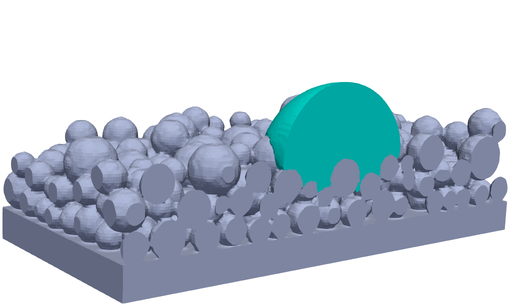}
}
\subfigure [$t = \SI{0.075}{\milli\second}$]
{
\includegraphics[width=0.23\textwidth]{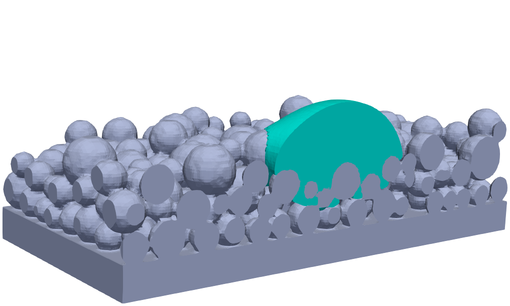}
}
\subfigure [$t = \SI{0.100}{\milli\second}$]
{
\includegraphics[width=0.23\textwidth]{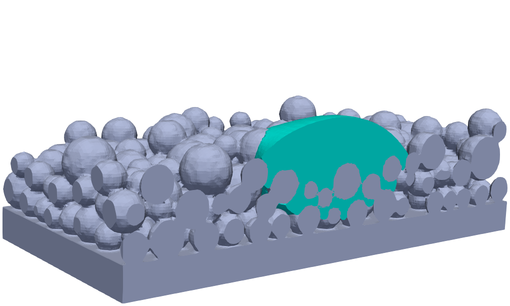}
}
\caption{Sectional view of BJT example with initial velocity $\SI{1}{\meter\per\second}$ of the liquid binder droplet: time series illustrating the impact of the liquid binder droplet on the powder bed.}
\label{fig:bjt_vel01_half}
\end{figure}

\begin{figure}[htbp]
\centering
\subfigure [$t = \SI{0.000}{\milli\second}$]
{
\includegraphics[width=0.23\textwidth]{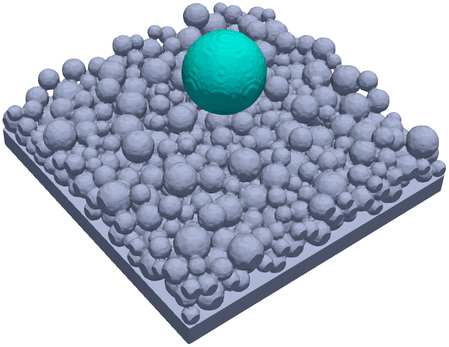}
}
\subfigure [$t = \SI{0.050}{\milli\second}$]
{
\includegraphics[width=0.23\textwidth]{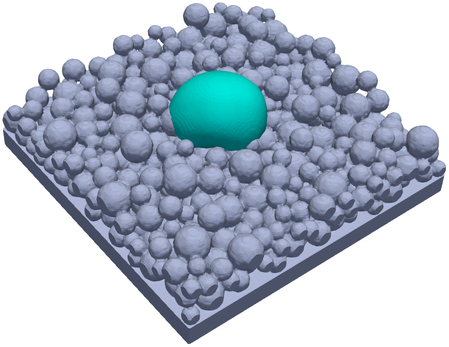}
}
\subfigure [$t = \SI{0.075}{\milli\second}$]
{
\includegraphics[width=0.23\textwidth]{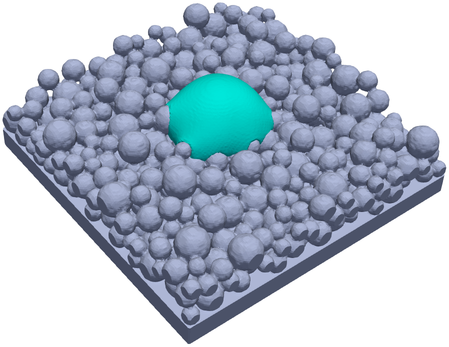}
}
\subfigure [$t = \SI{0.100}{\milli\second}$]
{
\includegraphics[width=0.23\textwidth]{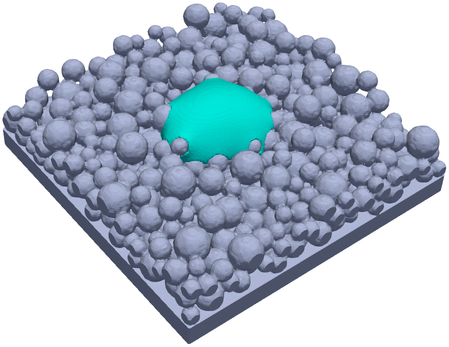}
}
\caption{Overall view of BJT example with initial velocity $\SI{1}{\meter\per\second}$ of the liquid binder droplet: time series illustrating the impact of the liquid binder droplet on the powder bed.}
\label{fig:bjt_vel01_full}
\end{figure}

\begin{figure}[htbp]
\centering
\subfigure [$t = \SI{0.000}{\milli\second}$]
{
\includegraphics[width=0.23\textwidth]{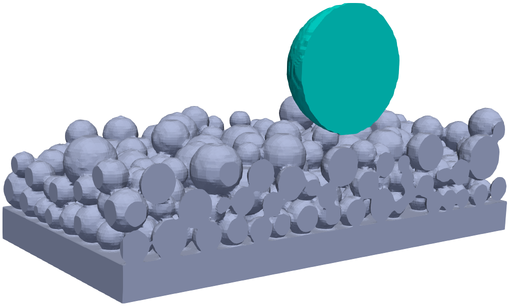}
}
\subfigure [$t = \SI{0.005}{\milli\second}$]
{
\includegraphics[width=0.23\textwidth]{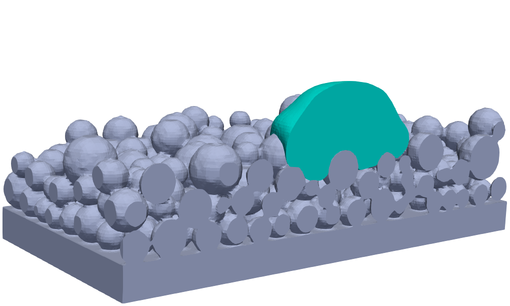}
}
\subfigure [$t = \SI{0.020}{\milli\second}$]
{
\includegraphics[width=0.23\textwidth]{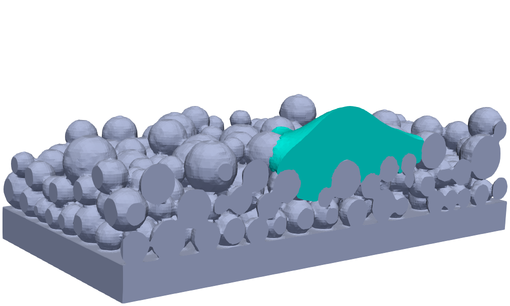}
}
\subfigure [$t = \SI{0.050}{\milli\second}$]
{
\includegraphics[width=0.23\textwidth]{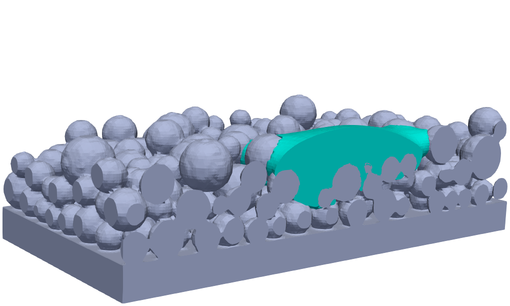}
}
\caption{Sectional view of BJT example with initial velocity $\SI{10}{\meter\per\second}$ of the liquid binder droplet: time series illustrating the impact of the liquid binder droplet on the powder bed.}
\label{fig:bjt_vel10_half}
\end{figure}

\begin{figure}[htbp]
\centering
\subfigure [$t = \SI{0.000}{\milli\second}$]
{
\includegraphics[width=0.23\textwidth]{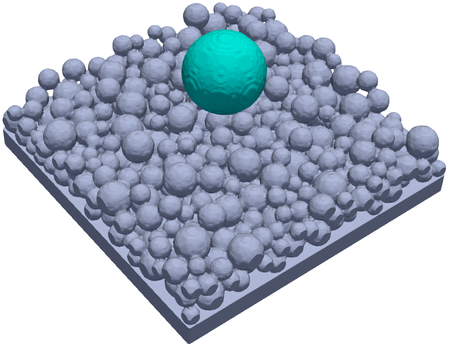}
}
\subfigure [$t = \SI{0.005}{\milli\second}$]
{
\includegraphics[width=0.23\textwidth]{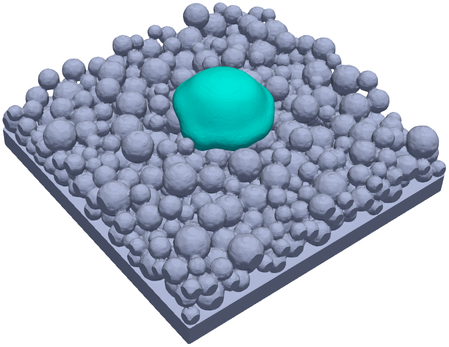}
}
\subfigure [$t = \SI{0.020}{\milli\second}$]
{
\includegraphics[width=0.23\textwidth]{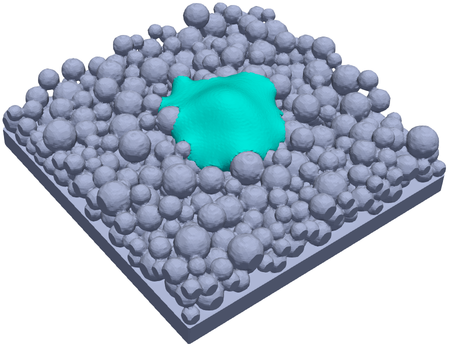}
}
\subfigure [$t = \SI{0.050}{\milli\second}$]
{
\includegraphics[width=0.23\textwidth]{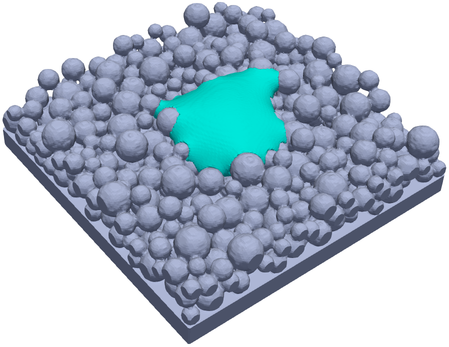}
}
\caption{Overall view of BJT example with initial velocity $\SI{10}{\meter\per\second}$ of the liquid binder droplet: time series illustrating the impact of the liquid binder droplet on the powder bed.}
\label{fig:bjt_vel10_full}
\end{figure}

\begin{figure}[htbp]
\centering
\subfigure [$t = \SI{0.000}{\milli\second}$]
{
\includegraphics[width=0.23\textwidth]{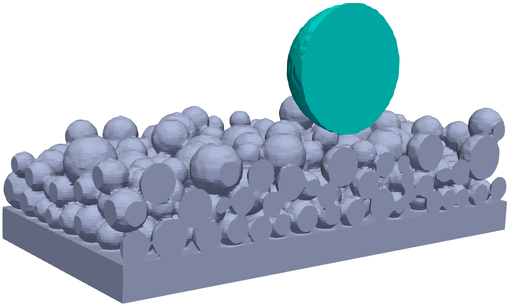}
}
\subfigure [$t = \SI{0.005}{\milli\second}$]
{
\includegraphics[width=0.23\textwidth]{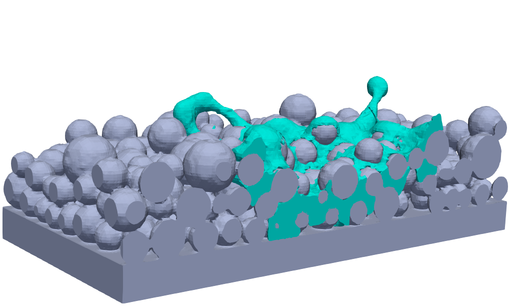}
}
\subfigure [$t = \SI{0.020}{\milli\second}$]
{
\includegraphics[width=0.23\textwidth]{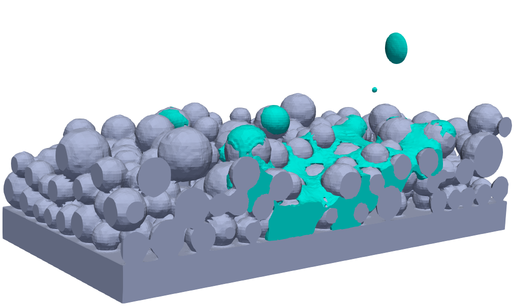}
}
\subfigure [$t = \SI{0.050}{\milli\second}$]
{
\includegraphics[width=0.23\textwidth]{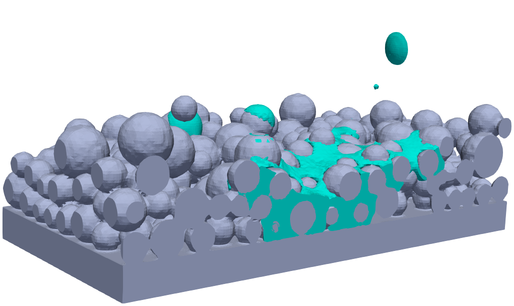}
}
\caption{Sectional view of BJT example with initial velocity $\SI{50}{\meter\per\second}$ of the liquid binder droplet: time series illustrating the impact of the liquid binder droplet on the powder bed.}
\label{fig:bjt_vel50_half}
\end{figure}

\begin{figure}[htbp]
\centering
\subfigure [$t = \SI{0.000}{\milli\second}$]
{
\includegraphics[width=0.23\textwidth]{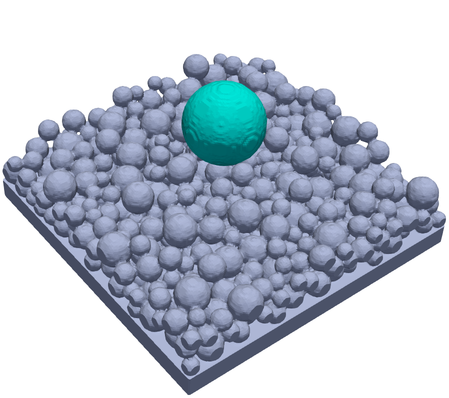}
}
\subfigure [$t = \SI{0.005}{\milli\second}$]
{
\includegraphics[width=0.23\textwidth]{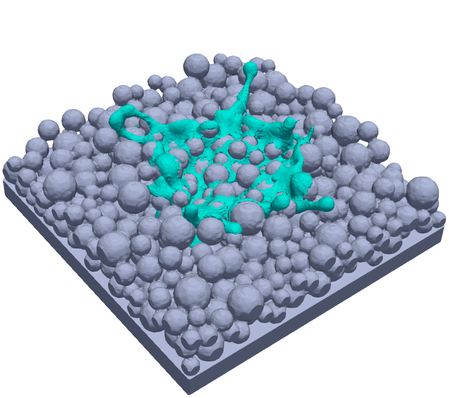}
}
\subfigure [$t = \SI{0.020}{\milli\second}$]
{
\includegraphics[width=0.23\textwidth]{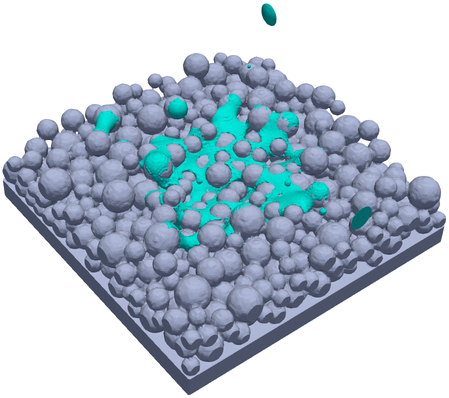}
}
\subfigure [$t = \SI{0.050}{\milli\second}$]
{
\includegraphics[width=0.23\textwidth]{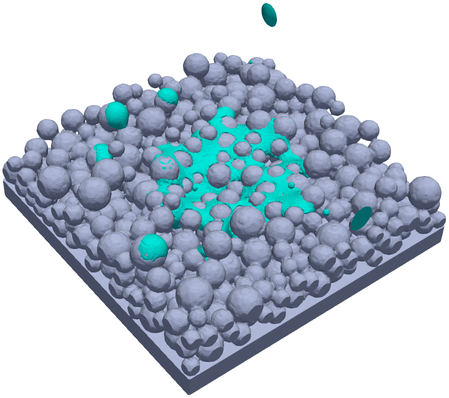}
}
\caption{Overall view of BJT example with initial velocity $\SI{50}{\meter\per\second}$ of the liquid binder droplet: time series illustrating the impact of the liquid binder droplet on the powder bed.}
\label{fig:bjt_vel50_full}
\end{figure}

In a next step, the initial velocity of the liquid binder droplet is increased to $\SI{10}{\meter\per\second}$. The obtained results are given in Figures~\ref{fig:bjt_vel10_half} and~\ref{fig:bjt_vel10_full} (see Supplementary Videos 3 and 4) until $t = \SI{0.05}{\milli\second}$. Due to its higher initial velocity, the liquid binder droplet is subject to oscillations upon impact with the powder bed. Besides, more pronounced powder dynamics are observed, i.e., the powder packing beneath the droplet is significantly compressed, and the liquid binder droplet is spread across a slightly larger area on the powder bed with less penetration depth as compared to the previous variant. At $t = \SI{0.05}{\milli\second}$ a slight downward movement of the binder
towards the substrate is still visible.

To enhance these effects, in the third variant the initial velocity of the liquid binder droplet is set to $\SI{50}{\meter\per\second}$. The obtained results are given in Figures~\ref{fig:bjt_vel50_half} and~\ref{fig:bjt_vel50_full} (see Supplementary Videos 5 and 6). The liquid binder droplet splashes into the powder bed, evoking highly dynamic motion of the metal powder particles. Besides, wetting of the liquid binder on the surface of individual powder particles can be observed. At the same time, spatters of liquid binder are ejected that subsequently adhere on the rigid walls surrounding the domain. The high impact velocity has the effect, that the packing density and uniformity of the powder bed is strongly disturbed and that the liquid binder is spread across a large area on the powder bed.

In sum, the results of these three variants demonstrate, that the microfluid dynamics of high-velocity binder droplets and their interaction with mobile powder particles can be captured by the proposed modeling framework in a robust manner. The replicated physical phenomena are typical for BJT processes, and, consequently, the proposed modeling framework can be recommended as a useful tool for detailed studies of these processes.

\subsection{Material jetting} \label{subsec:results_mjt}

In the next example, the behavior of molten metal droplets that are ejected on a substrate is examined. The setup is close to potential application scenarios in MJT processes. A domain with dimension $\SI{160}{\micro\meter} \times \SI{160}{\micro\meter} \times \SI{260}{\micro\meter}$ (discretized by approximately \num{1.7e6} SPH particles) is given. The lower part of the domain is occupied by a substrate of thickness $\SI{20}{\micro\meter}$. First, a spherical molten metal droplet with diameter $\SI{50}{\micro\meter}$ is placed centrally above the substrate with a distance of $\SI{40}{\micro\meter}$ with respect to the center of the drop. The remainder of the domain is filled with atmospheric gas and surrounded by rigid walls. The molten metal droplet is initialized with the temperature $\SI{2500}{\kelvin}$ and the velocity $\SI{1.6}{\meter\per\second}$ pointing in downward direction such that it is approaching the substrate.

\begin{figure}[htbp]
\centering
\subfigure [$t = \SI{0.025}{\milli\second}$]
{
\includegraphics[width=0.23\textwidth]{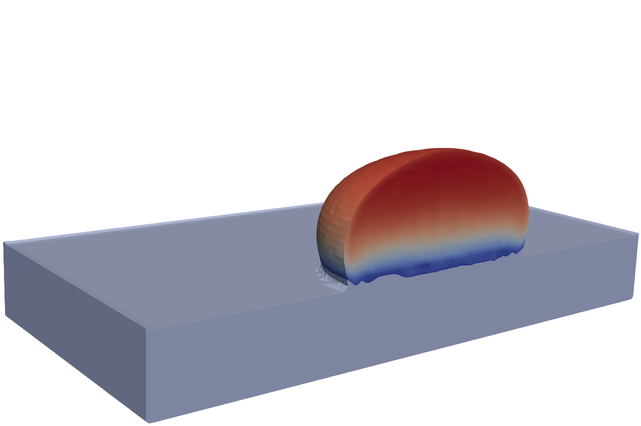}
}
\subfigure [$t = \SI{0.050}{\milli\second}$]
{
\includegraphics[width=0.23\textwidth]{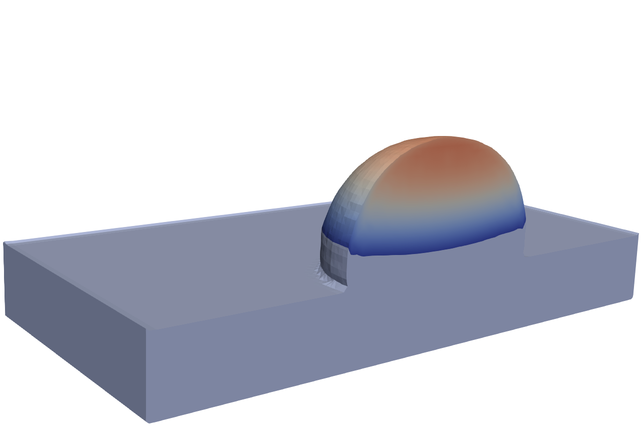}
}
\subfigure [$t = \SI{0.075}{\milli\second}$]
{
\includegraphics[width=0.23\textwidth]{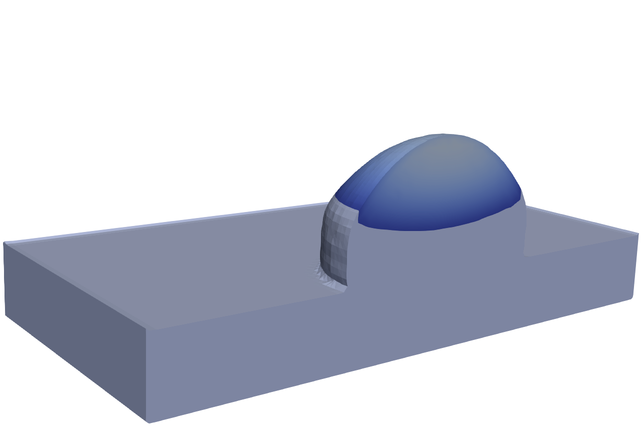}
}
\subfigure [$t = \SI{0.110}{\milli\second}$]
{
\includegraphics[width=0.23\textwidth]{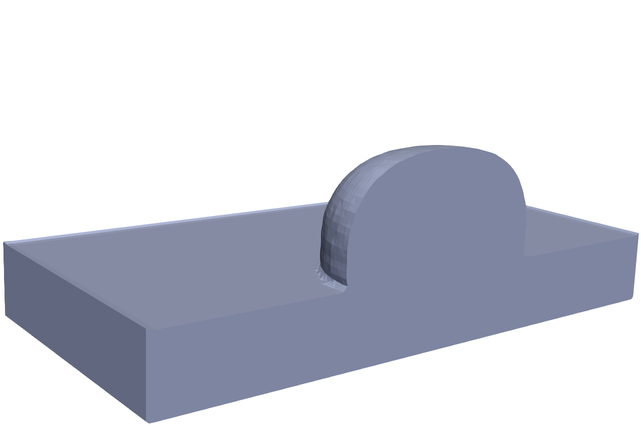}
}
\caption{Sectional view of MJT example with one molten metal droplet and initial temperature $\SI{500}{\kelvin}$ of the substrate: time series illustrating the impact of the molten metal droplet on the substrate and the solidification process with temperature field ranging from $\SI{1700}{\kelvin}$ (blue) to $\SI{2500}{\kelvin}$ (red).}
\label{fig:mjt_drop1_temp0500_half}
\end{figure}

\begin{figure}[htbp]
\centering
\subfigure [$t = \SI{0.025}{\milli\second}$]
{
\includegraphics[width=0.23\textwidth]{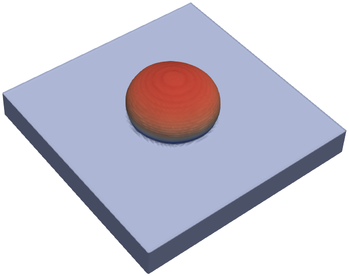}
}
\subfigure [$t = \SI{0.050}{\milli\second}$]
{
\includegraphics[width=0.23\textwidth]{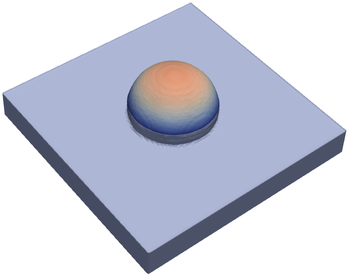}
}
\subfigure [$t = \SI{0.075}{\milli\second}$]
{
\includegraphics[width=0.23\textwidth]{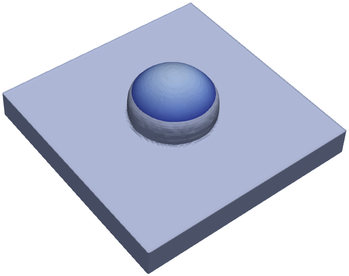}
}
\subfigure [$t = \SI{0.110}{\milli\second}$]
{
\includegraphics[width=0.23\textwidth]{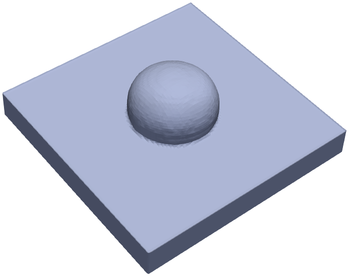}
}
\caption{Overall view of MJT example with one molten metal droplet and initial temperature $\SI{500}{\kelvin}$ of the substrate: time series illustrating the impact of the molten metal droplet on the substrate and the solidification process with temperature field ranging from $\SI{1700}{\kelvin}$ (blue) to $\SI{2500}{\kelvin}$ (red).}
\label{fig:mjt_drop1_temp0500_full}
\end{figure}

\begin{figure}[htbp]
\centering
\subfigure [$t = \SI{0.025}{\milli\second}$]
{
\includegraphics[width=0.23\textwidth]{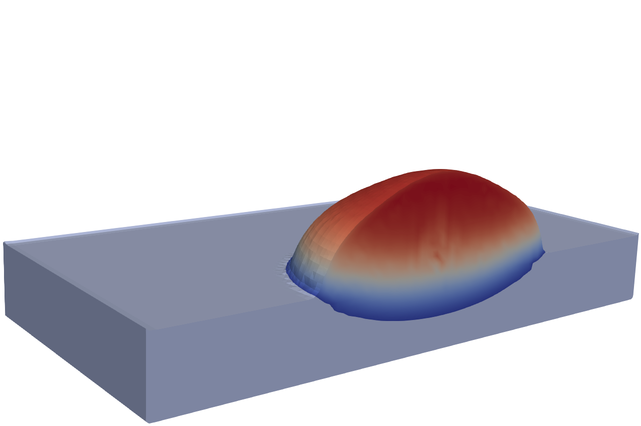}
}
\subfigure [$t = \SI{0.050}{\milli\second}$]
{
\includegraphics[width=0.23\textwidth]{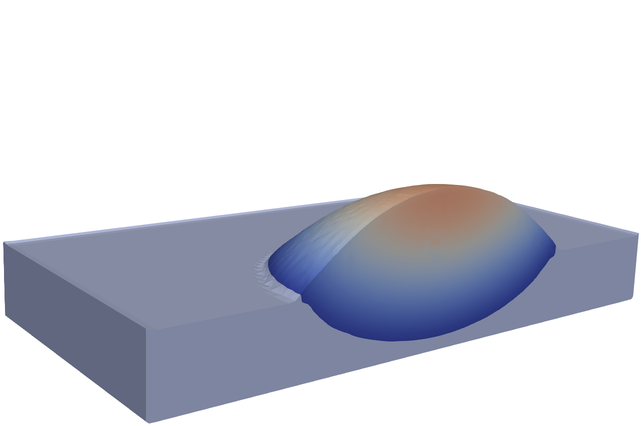}
}
\subfigure [$t = \SI{0.075}{\milli\second}$]
{
\includegraphics[width=0.23\textwidth]{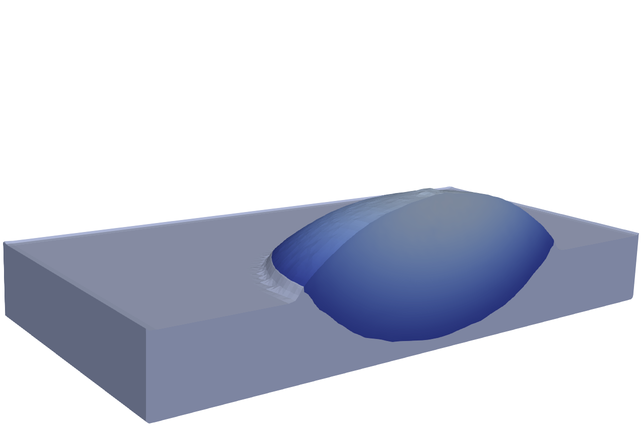}
}
\subfigure [$t = \SI{0.155}{\milli\second}$]
{
\includegraphics[width=0.23\textwidth]{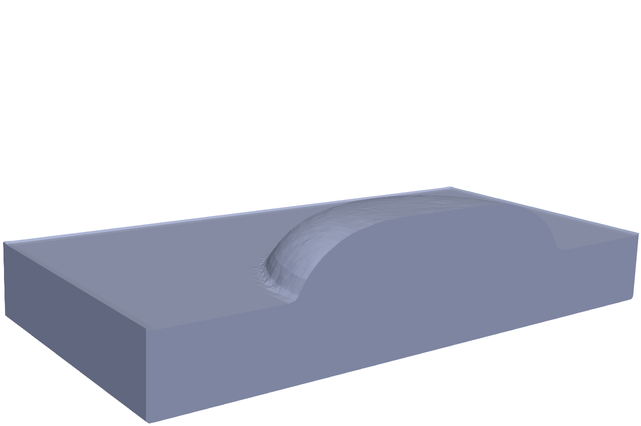}
}
\caption{Sectional view of MJT example with one molten metal droplet and initial temperature $\SI{1500}{\kelvin}$ of the substrate: time series illustrating the impact of the molten metal droplet on the substrate and the solidification process with temperature field ranging from $\SI{1700}{\kelvin}$ (blue) to $\SI{2500}{\kelvin}$ (red).}
\label{fig:mjt_drop1_temp1500_half}
\end{figure}

\begin{figure}[htbp]
\centering
\subfigure [$t = \SI{0.025}{\milli\second}$]
{
\includegraphics[width=0.23\textwidth]{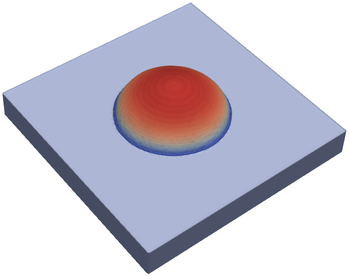}
}
\subfigure [$t = \SI{0.050}{\milli\second}$]
{
\includegraphics[width=0.23\textwidth]{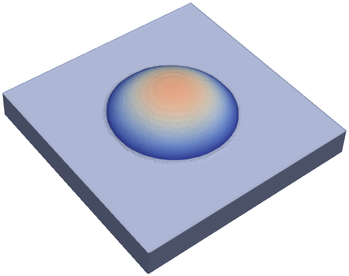}
}
\subfigure [$t = \SI{0.075}{\milli\second}$]
{
\includegraphics[width=0.23\textwidth]{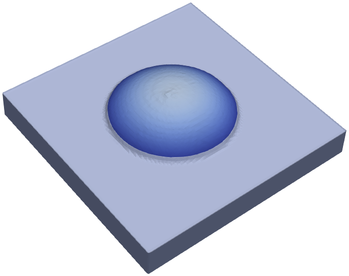}
}
\subfigure [$t = \SI{0.155}{\milli\second}$]
{
\includegraphics[width=0.23\textwidth]{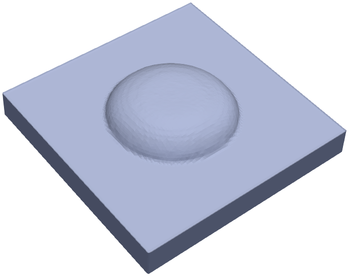}
}
\caption{Overall view of MJT example with one molten metal droplet and initial temperature $\SI{1500}{\kelvin}$ of the substrate: time series illustrating the impact of the molten metal droplet on the substrate and the solidification process with temperature field ranging from $\SI{1700}{\kelvin}$ (blue) to $\SI{2500}{\kelvin}$ (red).}
\label{fig:mjt_drop1_temp1500_full}
\end{figure}

A time series of the results obtained for a single molten metal droplet is given in Figures~\ref{fig:mjt_drop1_temp0500_half} and~\ref{fig:mjt_drop1_temp0500_full} (see Supplementary Videos 7 and 8). Typical surface tension-driven oscillations of the droplet can be observed caused by the impact with the substrate. The molten metal immediately solidifies when it comes into contact with the substrate. The substrate itself does not exhibit noticeable melting. In practice, this behavior could suggest insufficient or weak bonding to the substrate~\cite{Gilani2021, Simonelli2019}.

For the next variant, the initial temperature of the substrate is set to $\SI{1500}{\kelvin}$ to examine its influence on the resulting melt pool shape. The results obtained for this variant are illustrated in Figures~\ref{fig:mjt_drop1_temp1500_half} and~\ref{fig:mjt_drop1_temp1500_full} (see Supplementary Videos 9 and 10). A maximum melt pool depth of approximately $\SI{16}{\micro\meter}$ is observed at $t = \SI{0.075}{\milli\second}$, that can be explained with the higher initial temperature of the substrate compared to the previous variant. Accordingly, the oscillations of the droplet are less pronounced while the droplet itself is spread across a larger area on the substrate. As a result, the final solidified geometry is flatter than for the previous variant and a better bonding to the substrate can be expected due to the increased remelting depth.

In a next step, the example is extended to consider a total of two respectively three molten metal droplets. The droplets are each placed a distance of $\SI{40}{\micro\meter}$ and $\SI{200}{\micro\meter}$ respectively $\SI{40}{\micro\meter}$, $\SI{120}{\micro\meter}$, and $\SI{200}{\micro\meter}$ above the substrate. The lateral distance between the droplets is set to  $\SI{25}{\micro\meter}$. That is, the projection of the initial position of the droplets onto the substrate constitutes a line when considering two droplets, and an equilateral triangle when considering three droplets. The initial temperature of the substrate is set again to the standard value $\SI{500}{\kelvin}$, while the initial temperature of the molten metal droplets remains at $\SI{2500}{\kelvin}$. The obtained results for both variants are given in Figures~\ref{fig:mjt_drop2_temp0500_full} and~\ref{fig:mjt_drop3_temp0500_full} (see Supplementary Videos 11 and 12). For the two droplet variant, the first droplet solidifies in the same fashion as in the single droplet variant. When the droplet is almost completely solidified the second droplet comes into contact with the first one and shortly after with the substrate. Through contact with the second droplet, the interface of the first droplet partially remelts.  Due to contact with the substrate and the first droplet the second droplet solidifies slightly faster than observed for a single droplet. In the variant with three droplets, the second droplet hits the first droplet earlier such that less than half of the first droplet is solidified. The liquid and partially liquid droplets coalesce during the descend of the second droplet forming one large liquid phase. Due to the motion of the second droplet, the liquid exhibits large oscillations which are increased when the third droplet coalesces with the liquid phase. This oscillation leads to large scale ripples on the surface of the droplets during solidification. With respect to the practical realization of the process, the three droplet variant represents a case where the solidification time interval between successive droplet impacts is too short, leading to droplet-droplet coalescence, and in turn to increased surface roughness of the final part.

\begin{figure}[htbp]
\centering
\subfigure [$t = \SI{0.025}{\milli\second}$]
{
\includegraphics[width=0.233\textwidth]{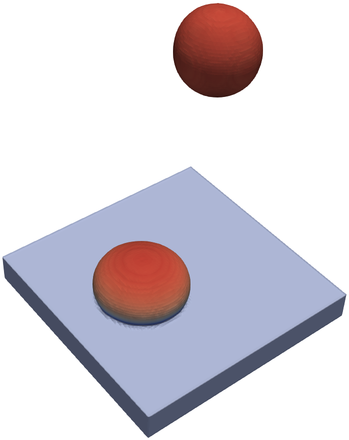}
}
\subfigure [$t = \SI{0.050}{\milli\second}$]
{
\includegraphics[width=0.230\textwidth]{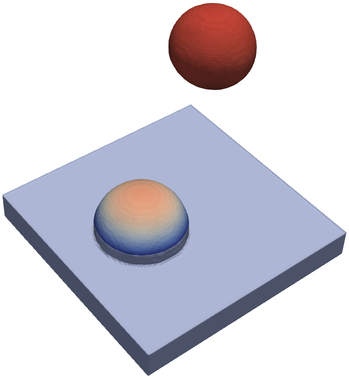}
}
\subfigure [$t = \SI{0.075}{\milli\second}$]
{
\includegraphics[width=0.230\textwidth]{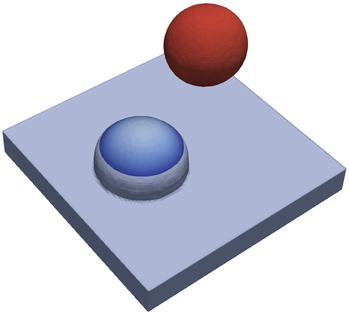}
}
\subfigure [$t = \SI{0.100}{\milli\second}$]
{
\includegraphics[width=0.230\textwidth]{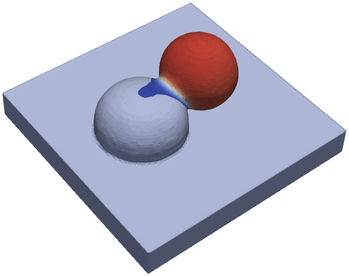}
}
\subfigure [$t = \SI{0.120}{\milli\second}$]
{
\includegraphics[width=0.230\textwidth]{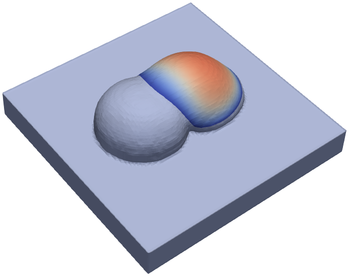}
}
\subfigure [$t = \SI{0.140}{\milli\second}$]
{
\includegraphics[width=0.230\textwidth]{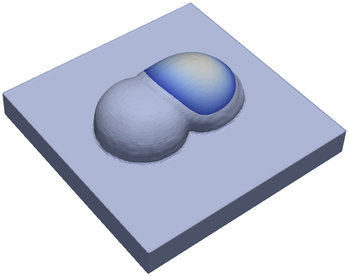}
}
\subfigure [$t = \SI{0.160}{\milli\second}$]
{
\includegraphics[width=0.230\textwidth]{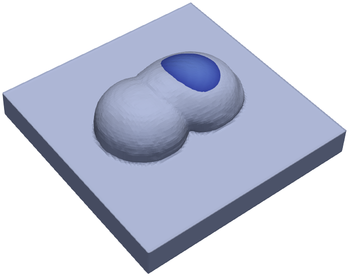}
}
\subfigure [$t = \SI{0.170}{\milli\second}$]
{
\includegraphics[width=0.230\textwidth]{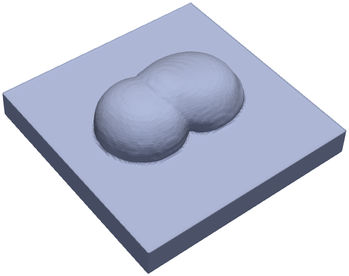}
}
\caption{Overall view of MJT example with two molten metal droplets and initial temperature $\SI{500}{\kelvin}$ of the substrate: time series illustrating the impact of the molten metal droplet on the substrate and the solidification process with temperature field ranging from $\SI{1700}{\kelvin}$ (blue) to $\SI{2500}{\kelvin}$ (red).}
\label{fig:mjt_drop2_temp0500_full}
\end{figure}

\begin{figure}[htbp]
\centering
\subfigure [$t = \SI{0.025}{\milli\second}$]
{
\includegraphics[width=0.233\textwidth]{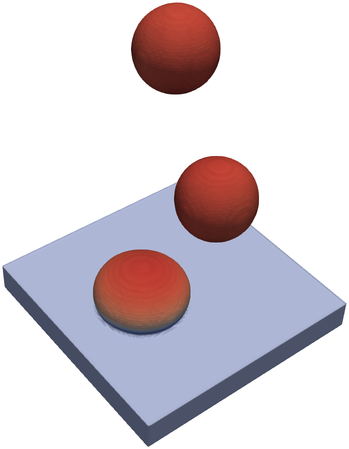}
}
\subfigure [$t = \SI{0.050}{\milli\second}$]
{
\includegraphics[width=0.23\textwidth]{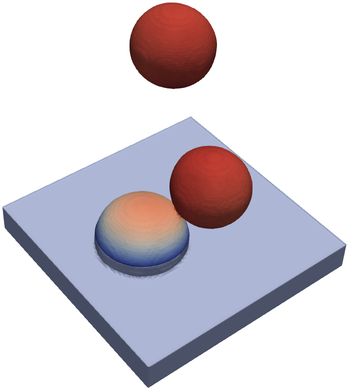}
}
\subfigure [$t = \SI{0.075}{\milli\second}$]
{
\includegraphics[width=0.23\textwidth]{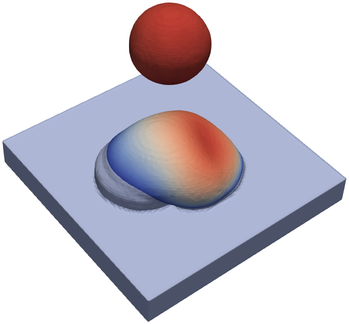}
}
\subfigure [$t = \SI{0.100}{\milli\second}$]
{
\includegraphics[width=0.23\textwidth]{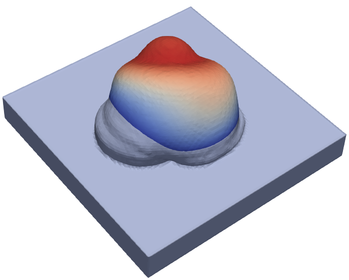}
}
\subfigure [$t = \SI{0.120}{\milli\second}$]
{
\includegraphics[width=0.23\textwidth]{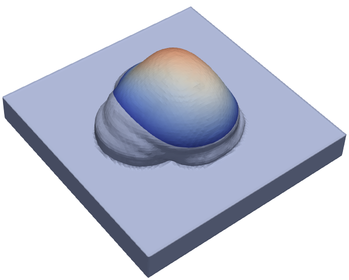}
}
\subfigure [$t = \SI{0.140}{\milli\second}$]
{
\includegraphics[width=0.23\textwidth]{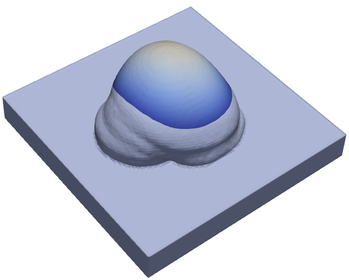}
}
\subfigure [$t = \SI{0.160}{\milli\second}$]
{
\includegraphics[width=0.23\textwidth]{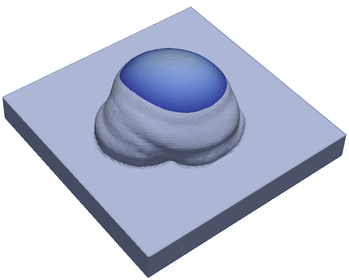}
}
\subfigure [$t = \SI{0.200}{\milli\second}$]
{
\includegraphics[width=0.23\textwidth]{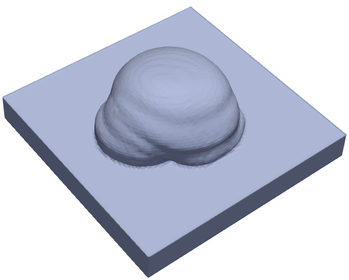}
}
\caption{Overall view of MJT example with three molten metal droplets and initial temperature $\SI{500}{\kelvin}$ of the substrate: time series illustrating the impact of the molten metal droplet on the substrate and the solidification process with temperature field ranging from $\SI{1700}{\kelvin}$ (blue) to $\SI{2500}{\kelvin}$ (red).}
\label{fig:mjt_drop3_temp0500_full}
\end{figure}

Altogether, the results of the four variants show the coalescence and dynamic solidification behavior of one or more molten metal droplets on a substrate, constituting typical effects of MJT processes. Above that, the modeling framework is able to capture the interplay of surface oscillations and rapid solidification leading to ripples on the interface as also observed in experiments.

\subsection{Directed energy deposition} \label{subsec:results_ded}

The setup of this example is motivated by a DED process, or more specifically an LPD process. Hence, consider a domain with dimension $\SI{260}{\micro\meter} \times \SI{260}{\micro\meter} \times \SI{240}{\micro\meter}$ (discretized by approximately \num{3.9e6} SPH particles). The lower part of the domain is occupied by a substrate of thickness $\SI{20}{\micro\meter}$. A total of five individual powder particles each with diameter $\SI{16}{\micro\meter}$ are aligned on an axis with inclination angle $\SI{45}{\degree}$ through the center of the substrate, simulating the powder stream. Initially, the centers of the individual powder particles are at a distance of $\SI{40}{\micro\meter}$, $\SI{60}{\micro\meter}$, $\SI{80}{\micro\meter}$, $\SI{100}{\micro\meter}$, and $\SI{120}{\micro\meter}$ above the substrate. The remainder of the domain is filled with atmospheric gas and surrounded by rigid walls. The powder particles are all initialized with the velocity $\SI{10}{\meter\per\second}$ towards the center of the substrate. A laser beam with total power $P_{l} = \SI{312}{\watt}$ and effective diameter $d_{w} = \SI{140}{\micro\meter}$ is acting in downward direction at the center of the domain. The laser melts the substrate and to some extend also the injected powder. Two variants are considered in the following: First, the laser remains continuously switched on. Second, the laser is switched off after $t = \SI{0.012}{\milli\second}$ to study solidification.

\begin{figure}[htbp]
\centering
\subfigure [$t = \SI{0.000}{\milli\second}$]
{
\includegraphics[width=0.23\textwidth]{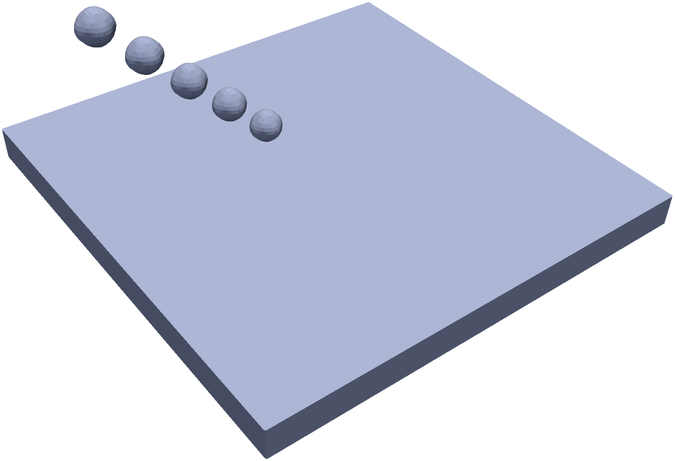}
}
\subfigure [$t = \SI{0.004}{\milli\second}$]
{
\includegraphics[width=0.23\textwidth]{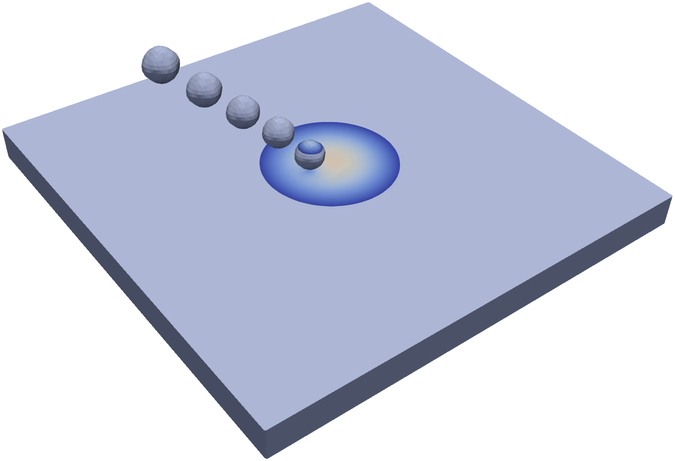}
}
\subfigure [$t = \SI{0.006}{\milli\second}$]
{
\includegraphics[width=0.23\textwidth]{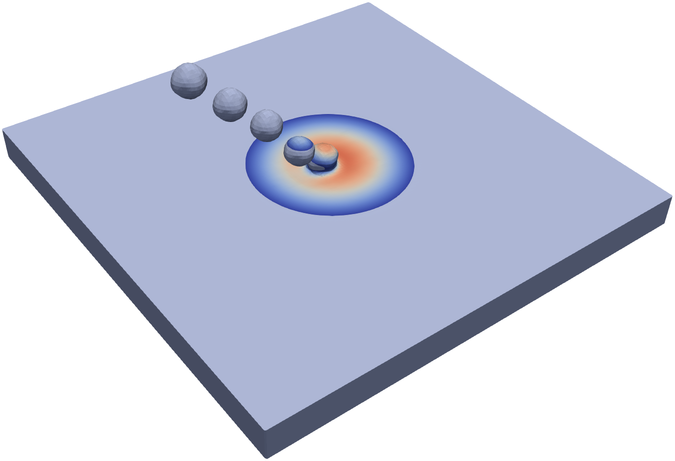}
}
\subfigure [$t = \SI{0.008}{\milli\second}$]
{
\includegraphics[width=0.23\textwidth]{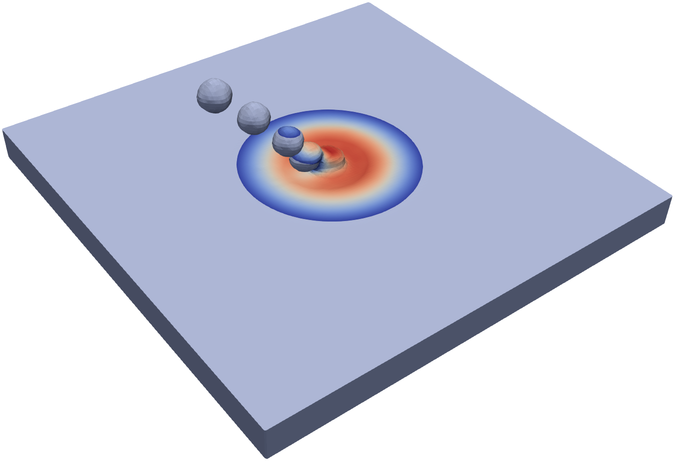}
}
\subfigure [$t = \SI{0.009}{\milli\second}$]
{
\includegraphics[width=0.23\textwidth]{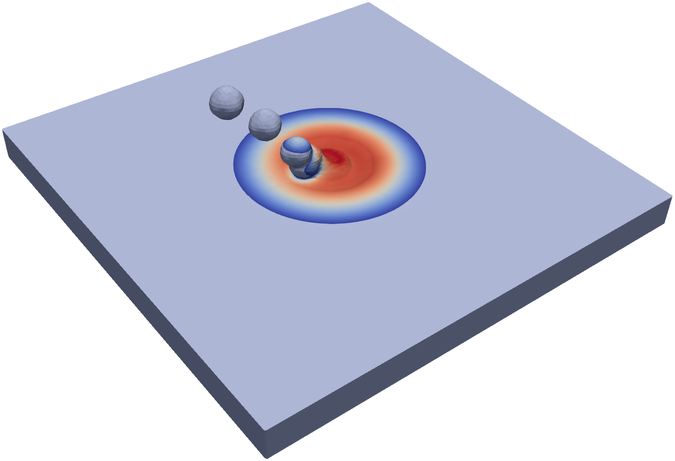}
}
\subfigure [$t = \SI{0.010}{\milli\second}$]
{
\includegraphics[width=0.23\textwidth]{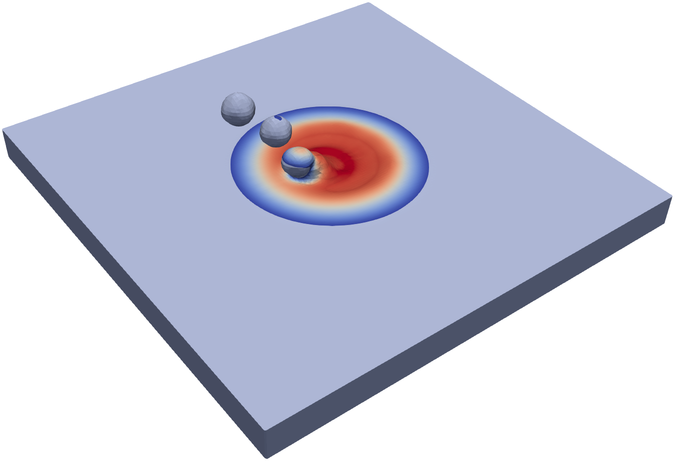}
}
\subfigure [$t = \SI{0.011}{\milli\second}$]
{
\includegraphics[width=0.23\textwidth]{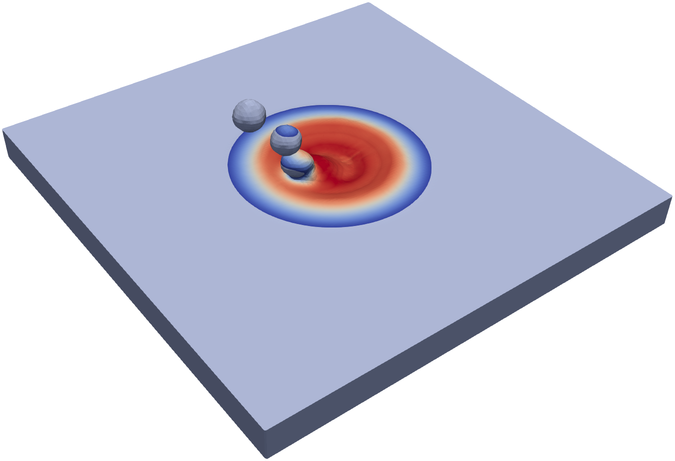}
}
\subfigure [$t = \SI{0.012}{\milli\second}$]
{
\includegraphics[width=0.23\textwidth]{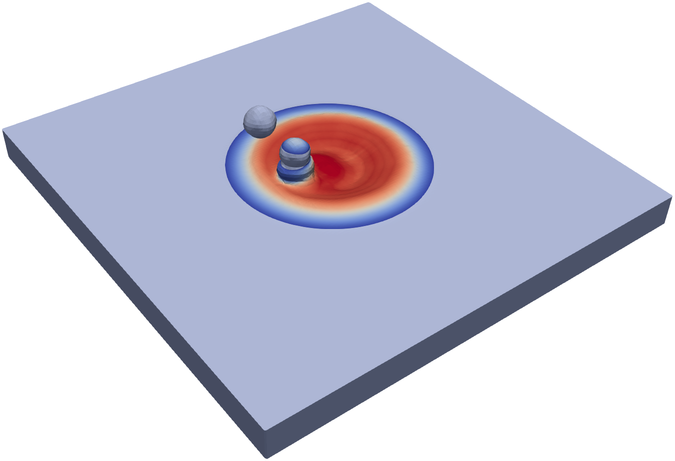}
}
\subfigure [$t = \SI{0.013}{\milli\second}$]
{
\includegraphics[width=0.23\textwidth]{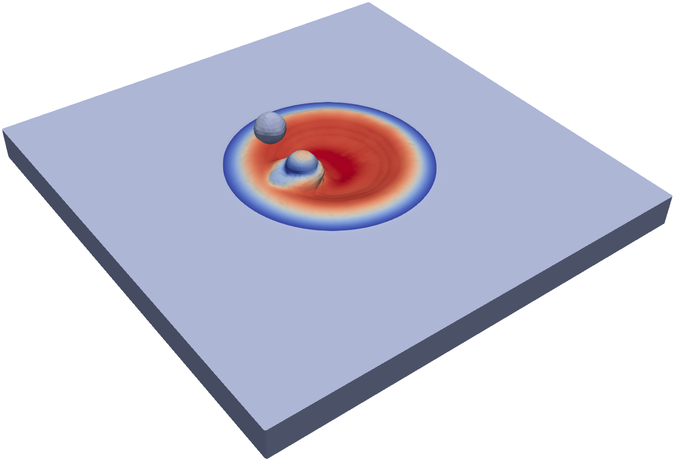}
}
\subfigure [$t = \SI{0.014}{\milli\second}$]
{
\includegraphics[width=0.23\textwidth]{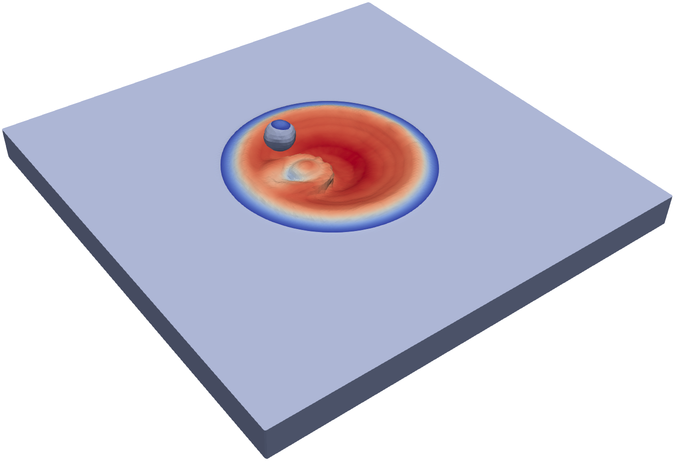}
}
\subfigure [$t = \SI{0.016}{\milli\second}$]
{
\includegraphics[width=0.23\textwidth]{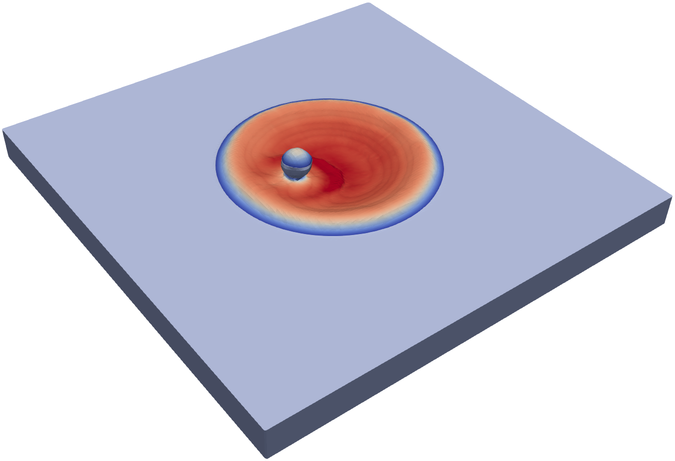}
}
\subfigure [$t = \SI{0.018}{\milli\second}$]
{
\includegraphics[width=0.23\textwidth]{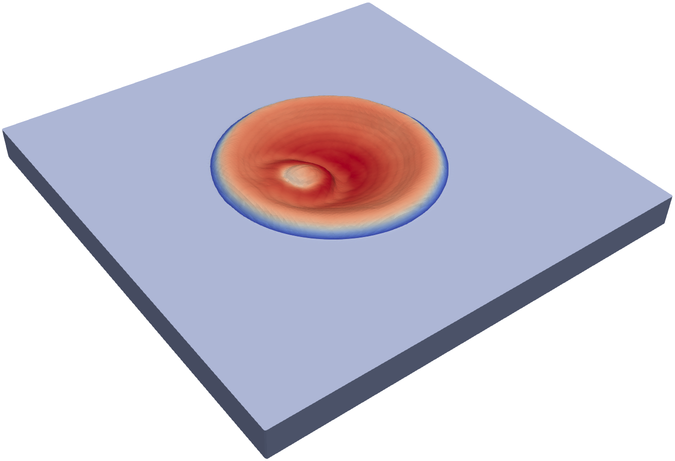}
}
\caption{Overall view of DED example with laser continuously switched on: time series illustrating the powder dynamics and the melt pool shape with temperature field ranging from $\SI{1700}{\kelvin}$ (blue) to $\SI{3400}{\kelvin}$ (red).}
\label{fig:ded_laser_continuous_on}
\end{figure}

In the first variant (see Figure~\ref{fig:ded_laser_continuous_on} and Supplementary Video 13) a melt pool starts to form due to heating by the laser. As the first powder particle comes in the vicinity of the laser beam at $t\approx\SI{0.004}{\milli\second}$, the top of the powder particle slowly starts to melt, but enters the melt pool still in solid form at $t\approx\SI{0.006}{\milli\second}$. Though, a deflection of the powder particle can be observed as it hits the solid substrate at the bottom of the melt pool which is still shallow at the beginning. Soon thereafter, at $t\approx\SI{0.007}{\milli\second}$, the second powder particle collides with the first one. When the third powder particle hits the second and immerses in its liquid phase ($t\approx\SI{0.009}{\milli\second}$), the first powder particle is almost fully liquefied. At $t\approx\SI{0.012}{\milli\second}$ the melt pool grew large enough such that solid powder particles can immerse in the melt pool. Accordingly, the powder particles melt through heat conduction from the surrounding melt pool as typical for LPD. During the simulation the melt pool increases in size to a final diameter of approximately $\SI{130}{\micro\meter}$ at $t\approx\SI{0.018}{\milli\second}$. Note that partially molten powder particles are deflected from their straight trajectory by recoil pressure forces which emerges when the peak temperature on the surface exceeds the boiling temperature. Hence, the powder particles do not hit the melt pool in its center. For the practical realization, this evaporation-induced deflection can be taken into account by adapting the orientation of the powder stream.

\begin{figure}[htbp]
\centering
\subfigure [$t = \SI{0.013}{\milli\second}$]
{
\includegraphics[width=0.23\textwidth]{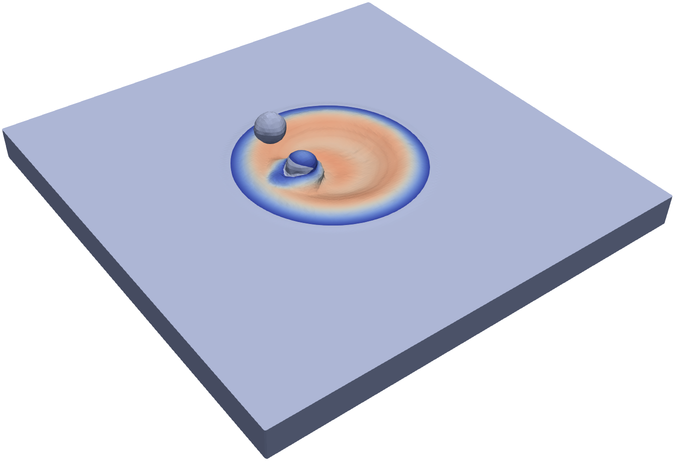}
}
\subfigure [$t = \SI{0.014}{\milli\second}$]
{
\includegraphics[width=0.23\textwidth]{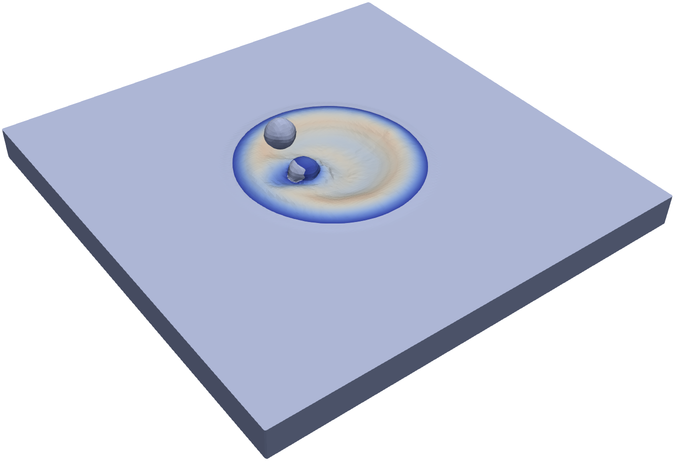}
}
\subfigure [$t = \SI{0.016}{\milli\second}$]
{
\includegraphics[width=0.23\textwidth]{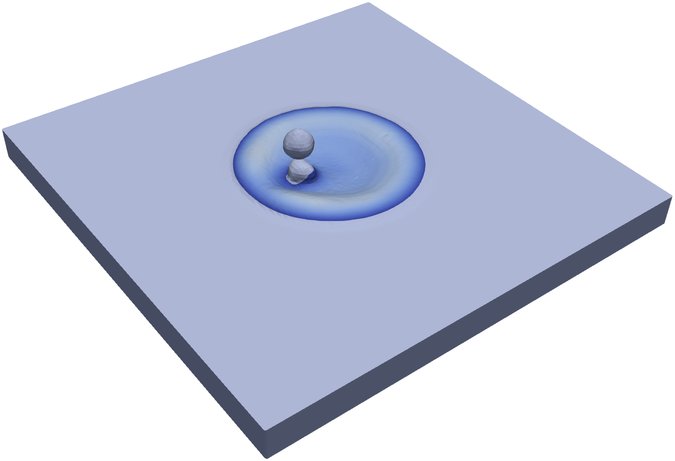}
}
\subfigure [$t = \SI{0.018}{\milli\second}$]
{
\includegraphics[width=0.23\textwidth]{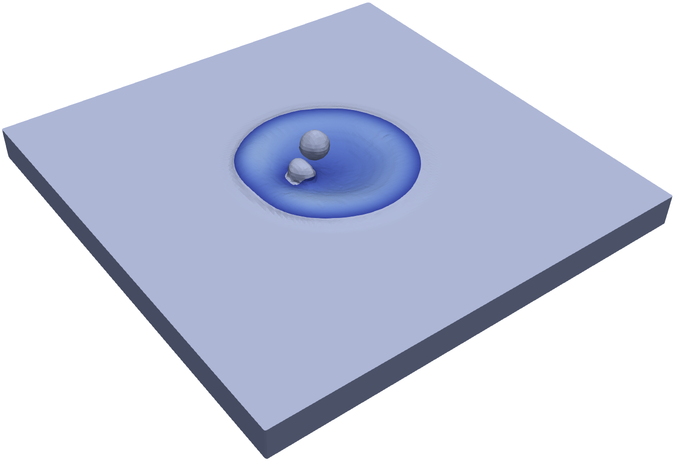}
}
\subfigure [$t = \SI{0.020}{\milli\second}$]
{
\includegraphics[width=0.23\textwidth]{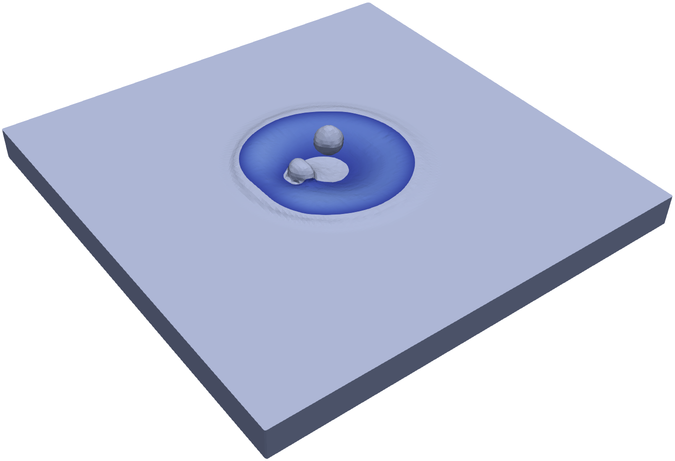}
}
\subfigure [$t = \SI{0.022}{\milli\second}$]
{
\includegraphics[width=0.23\textwidth]{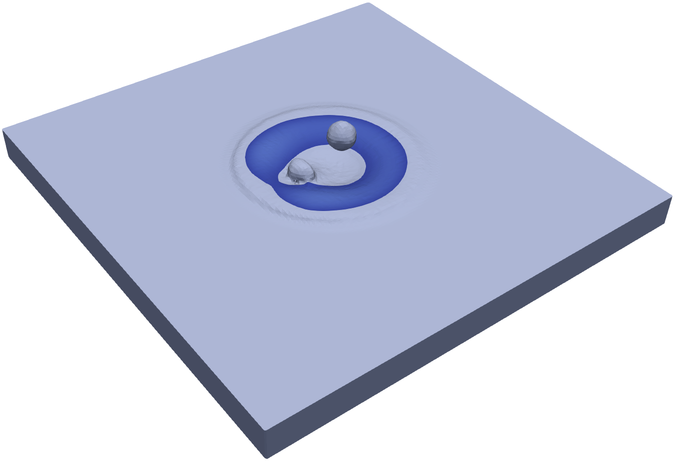}
}
\subfigure [$t = \SI{0.024}{\milli\second}$]
{
\includegraphics[width=0.23\textwidth]{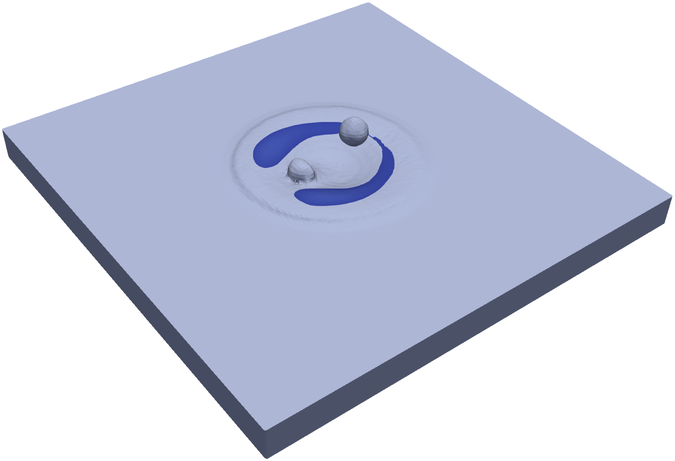}
}
\subfigure [$t = \SI{0.026}{\milli\second}$]
{
\includegraphics[width=0.23\textwidth]{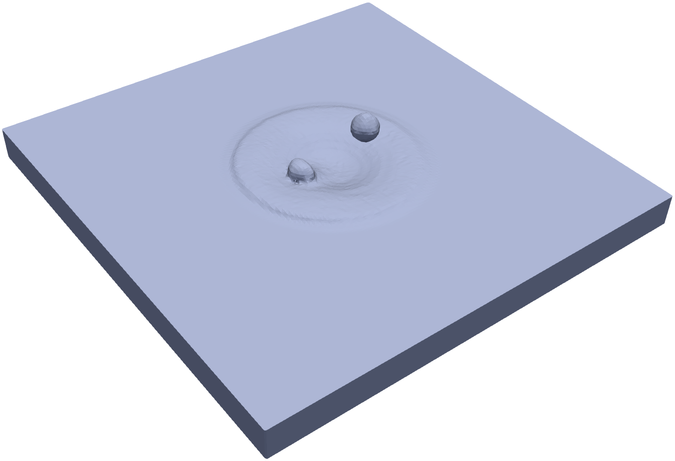}
}
\caption{Overall view of DED example with laser switched off at $t = \SI{0.012}{\milli\second}$: time series illustrating the powder dynamics and the melt pool shape with temperature field ranging from $\SI{1700}{\kelvin}$ (blue) to $\SI{3400}{\kelvin}$ (red).}
\label{fig:ded_laser_switched_off}
\end{figure}

In the second variant the laser is switched off at $t = \SI{0.012}{\milli\second}$ such that the same observations as in the previous example can be made until $t = \SI{0.012}{\milli\second}$ (see Figure~\ref{fig:ded_laser_continuous_on}). Figure~\ref{fig:ded_laser_switched_off} shows the solidification process (see Supplementary Video 14). As soon as the laser is switched off the melt pool rapidly cools down and gradually solidifies. The powder particle coming into contact with the melt pool at the time of the switch-off only partially melts and rapidly solidifies such that the last powder particle bounces off and is deflected, finally flying above the substrate at $t\approx\SI{0.026}{\milli\second}$.

With five powder particles only a small part of a realistic LPD process is considered in this example. Still, both variants allow to study in detail the interaction of mobile powder particles with the laser heat source, the molten metal, and other powder particles, thus capturing important aspects of LPD. Therefore, the proposed modeling framework is deemed suitable to model this process.

\subsection{Powder bed fusion} \label{subsec:results_pbf}

This example examines an SLM process, a variant of PBF, utilizing a laser beam as local heat source. That is, consider a domain with dimension $\SI{320}{\micro\meter} \times \SI{320}{\micro\meter} \times \SI{240}{\micro\meter}$ (discretized by approximately \num{5.9e6} SPH particles). The lower part of the domain is occupied by a substrate of thickness $\SI{80}{\micro\meter}$. A powder bed consisting of a total of 500 individual powder particles with diameters between $\SI{16}{\micro\meter}$ and $\SI{32}{\micro\meter}$ is resting on the substrate. The remainder of the domain is filled with atmospheric gas and surrounded by rigid walls. A spatially fixed laser beam with total power $P_{l} = \SI{312}{\watt}$ and effective diameter $d_{w} = \SI{140}{\micro\meter}$ is acting in downward direction in the center of the domain melting the powder and the substrate until $t = \SI{0.1}{\milli\second}$. Subsequently, the laser is switched off to allow for solidification until $t = \SI{0.25}{\milli\second}$. In this example, the time step size is set to~$\Delta{}t = \SI{0.5e-6}{\milli\second}$.

A time series of the obtained results is given in Figures~\ref{fig:pbf_pointmelting_half} and~\ref{fig:pbf_pointmelting_full} (see Supplementary Videos 15 and 16). The laser beam locally melts the granular metal powder such that under the effect of surface tension a melt pool with smooth surface is formed. When the peak temperature in the melt pool center exceeds the boiling temperature, recoil pressure forces induce the formation of a deep keyhole-shaped depression of the melt pool. In particular, recoil pressure-induced waves can be observed, which propagate from the top to the bottom of the keyhole while increasing in their magnitude. These wave patterns are well-known for laser-metal interactions in the high power regime, and their instable growth behavior can be traced back to the strong mutual coupling of local keyhole surface curvature, laser energy absorption and recoil pressure \cite{Kouraytem2019}. Due to the recoil pressure forces (and thereby induced gas flow), powder particles are dynamically ejected from the edge of the melt pool. Some of the ejected powder particles are partially molten and resolidify in the atmospheric gas. Boundary effects are visible as some of the ejected powder particles bounce off the rigid walls (not visualized) surrounding the computational domain. After the laser is switched off at $t = \SI{0.1}{\milli\second}$, the bottom of the melt pool solidifies very quickly, i.e., faster than the surface tension can smoothen out the melt pool depression, leaving a deep indentation. The larger portion of the molten metal at the top of the melt pool cools down slower while constricting under the action of surface tension to form a pore.

\begin{figure}[htbp]
\centering
\subfigure [$t = \SI{0.000}{\milli\second}$]
{
\includegraphics[width=0.23\textwidth]{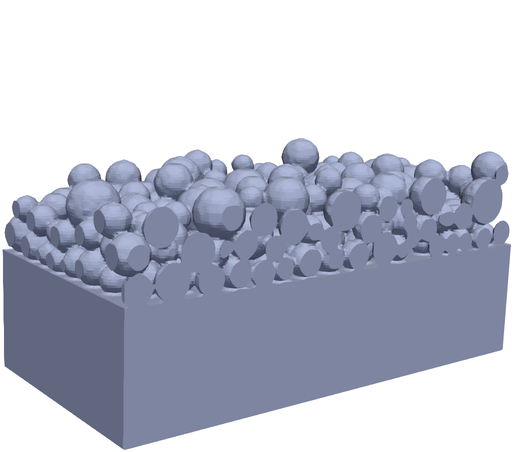}
}
\subfigure [$t = \SI{0.050}{\milli\second}$]
{
\includegraphics[width=0.23\textwidth]{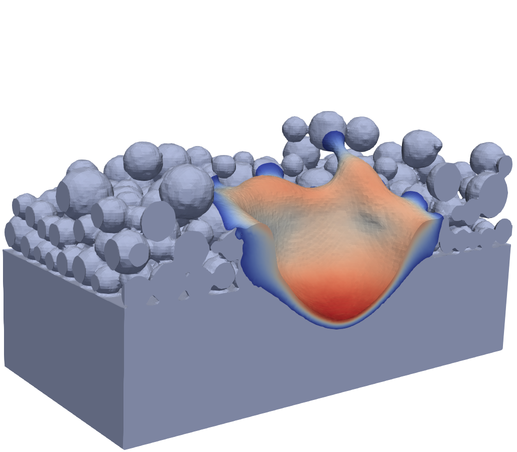}
}
\subfigure [$t = \SI{0.075}{\milli\second}$]
{
\includegraphics[width=0.23\textwidth]{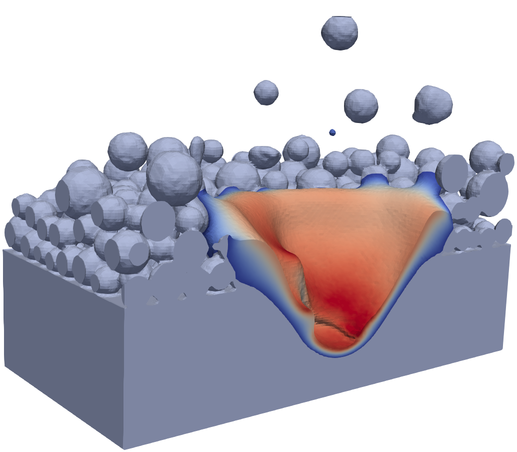}
}
\subfigure [$t = \SI{0.100}{\milli\second}$]
{
\includegraphics[width=0.23\textwidth]{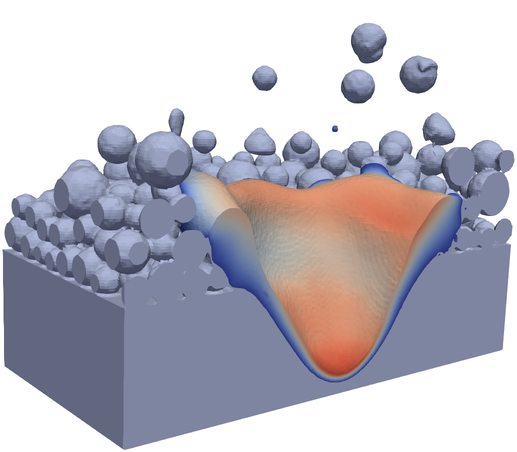}
}
\subfigure [$t = \SI{0.125}{\milli\second}$]
{
\includegraphics[width=0.23\textwidth]{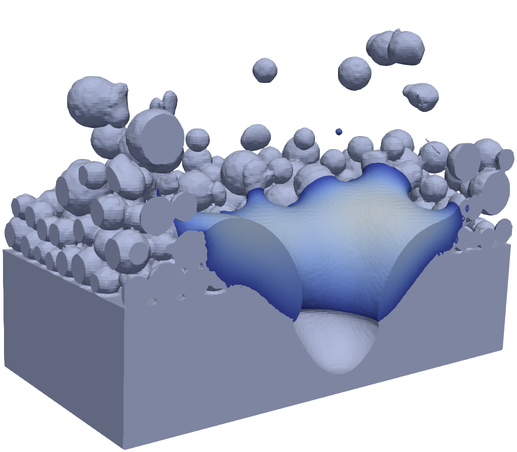}
}
\subfigure [$t = \SI{0.150}{\milli\second}$]
{
\includegraphics[width=0.23\textwidth]{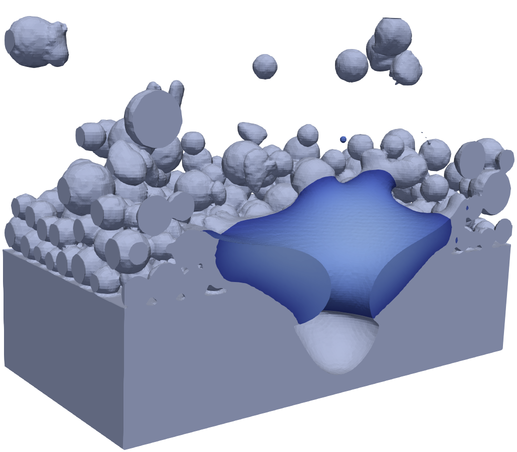}
}
\subfigure [$t = \SI{0.200}{\milli\second}$]
{
\includegraphics[width=0.23\textwidth]{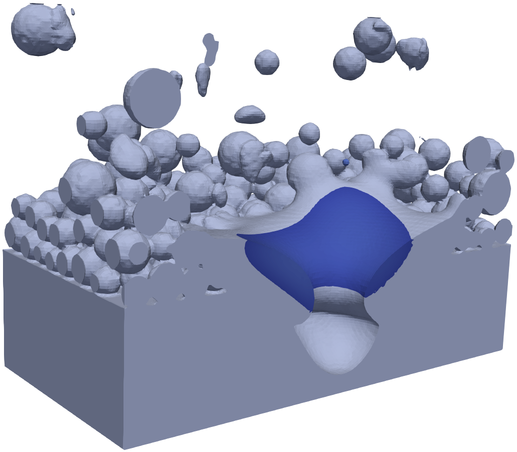}
}
\subfigure [$t = \SI{0.250}{\milli\second}$]
{
\includegraphics[width=0.23\textwidth]{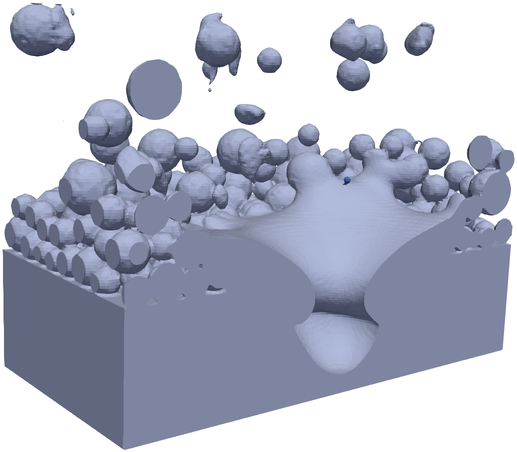}
}
\caption{Sectional view of point melting PBF example: time series illustrating the powder dynamics and the melt pool shape with temperature field ranging from $\SI{1700}{\kelvin}$ (blue) to $\SI{3400}{\kelvin}$ (red).}
\label{fig:pbf_pointmelting_half}
\end{figure}

\begin{figure}[htbp]
\centering
\subfigure [$t = \SI{0.000}{\milli\second}$]
{
\includegraphics[width=0.23\textwidth]{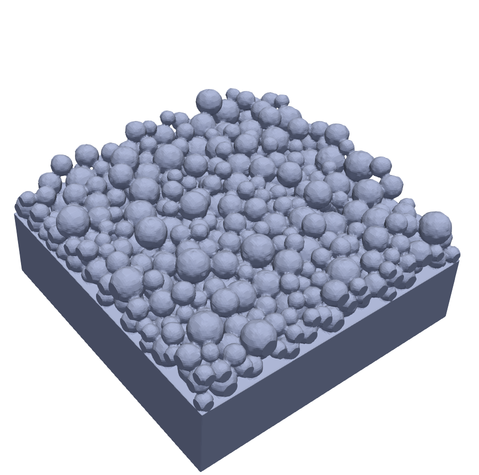}
}
\subfigure [$t = \SI{0.050}{\milli\second}$]
{
\includegraphics[width=0.23\textwidth]{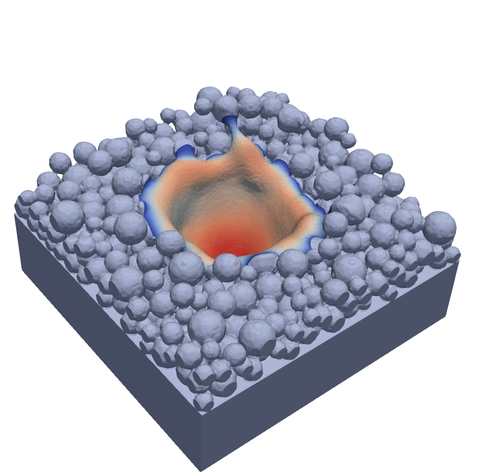}
}
\subfigure [$t = \SI{0.075}{\milli\second}$]
{
\includegraphics[width=0.23\textwidth]{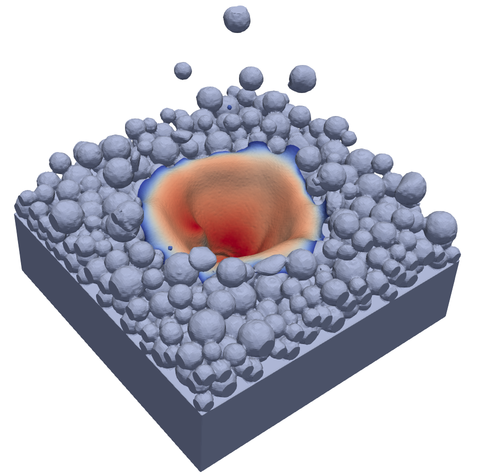}
}
\subfigure [$t = \SI{0.100}{\milli\second}$]
{
\includegraphics[width=0.23\textwidth]{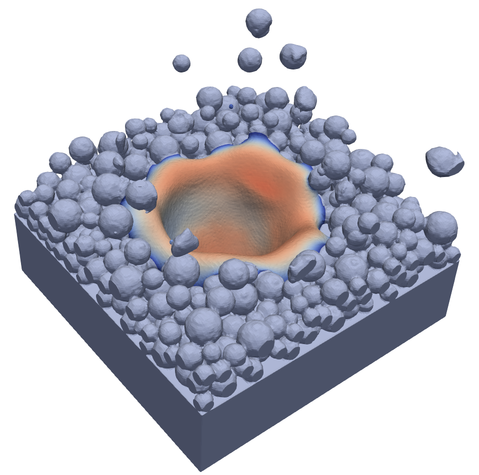}
}
\subfigure [$t = \SI{0.125}{\milli\second}$]
{
\includegraphics[width=0.23\textwidth]{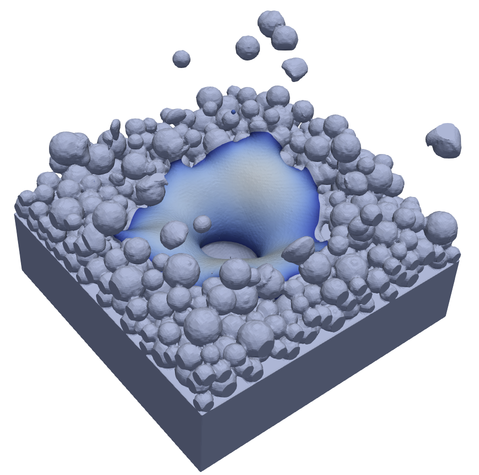}
}
\subfigure [$t = \SI{0.150}{\milli\second}$]
{
\includegraphics[width=0.23\textwidth]{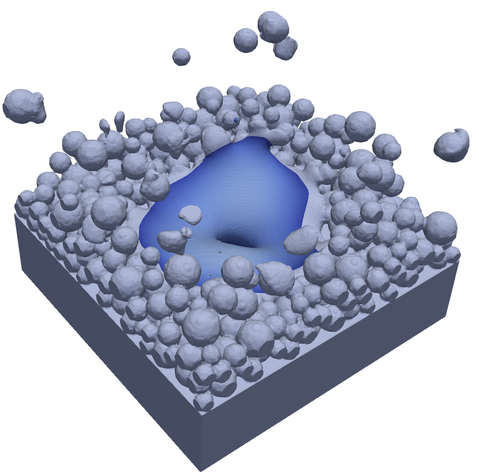}
}
\subfigure [$t = \SI{0.200}{\milli\second}$]
{
\includegraphics[width=0.23\textwidth]{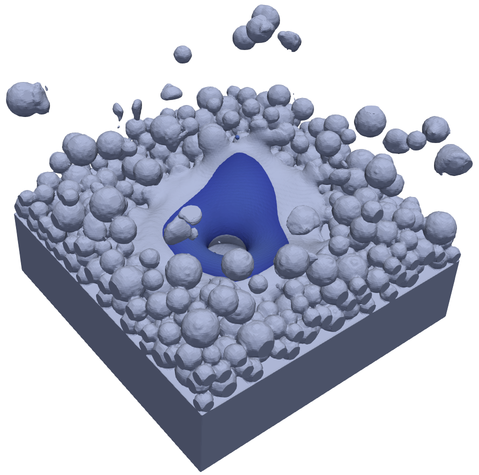}
}
\subfigure [$t = \SI{0.250}{\milli\second}$]
{
\includegraphics[width=0.23\textwidth]{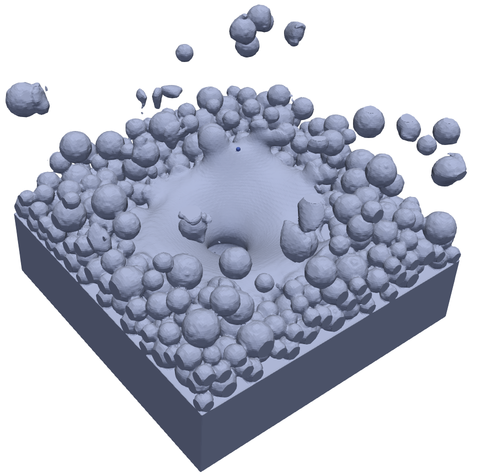}
}
\caption{Overall view of point melting PBF example: time series illustrating the powder dynamics and the melt pool shape with temperature field ranging from $\SI{1700}{\kelvin}$ (blue) to $\SI{3400}{\kelvin}$ (red).}
\label{fig:pbf_pointmelting_full}
\end{figure}

In a next step, the domain is extended by a factor of three in longitudinal direction to examine the case of line melting. Hence, the dimension of the domain covers $\SI{960}{\micro\meter} \times \SI{320}{\micro\meter} \times \SI{240}{\micro\meter}$ (discretized by approximately \num{17.3e6} SPH particles) and the powder bed is constituted by a total of 1500 individual powder particles. The laser beam is switched on at a distance of $\SI{160}{\micro\meter}$ from one end of the domain and remains at rest until $t = \SI{0.05}{\milli\second}$. Then the laser beam starts moving with the constant velocity $\SI{2}{\meter\per\second}$ in longitudinal direction towards the other end of the domain, crossing a total distance of $\SI{640}{\micro\meter}$, until it stops at $t = \SI{0.37}{\milli\second}$. At the same time, the laser is switched off to allow for solidification. Figures~\ref{fig:pbf_linemelting_half} and~\ref{fig:pbf_linemelting_full} (see Supplementary Videos 17 and 18) show a time series of the obtained results. In the initial phase, similar observations as for the point melting variant can be made. As soon as the laser starts its longitudinal movement, the propagating melt front compresses and pushes ahead the powder feedstock in front of the laser. Due to this effect, which has to the best of the authors' knowledge not been observed in previous studies, the initial packing structure of powder material ahead of the laser beam is strongly distorted before melting takes place. In addition, recoil pressure forces (and thereby induced gas flow) lead to dynamic particle ejections and instable wave patterns across the front melt pool wall (where temperatures are highest) as already observed in the point melting case above. Eventually, a continuous melt track lags behind the laser and gradually starts to solidify, leaving a small island of molten metal behind. After the laser is switched off, the thin film of melt directly beneath the laser beam center cools down very quickly, leaving back a surface indentation after solidification.

\begin{figure}[htbp]
\centering
\subfigure [$t = \SI{0.000}{\milli\second}$]
{
\includegraphics[width=0.31\textwidth]{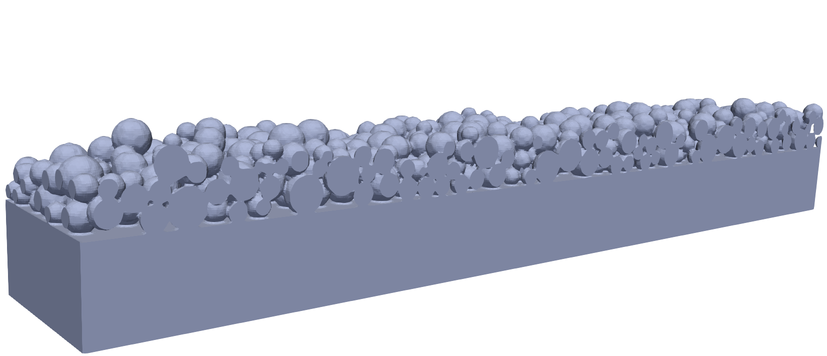}
}
\subfigure [$t = \SI{0.050}{\milli\second}$]
{
\includegraphics[width=0.31\textwidth]{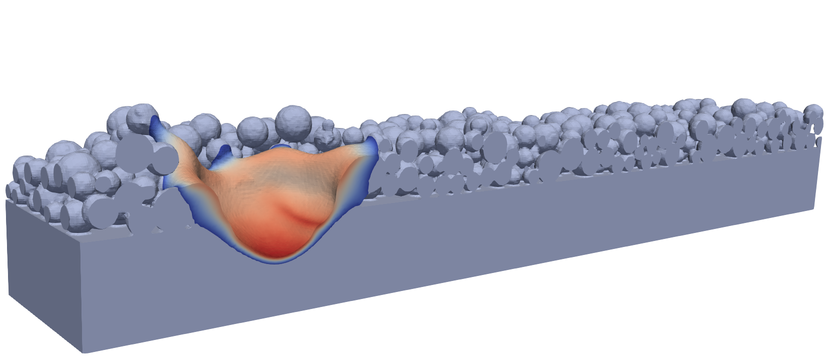}
}
\subfigure [$t = \SI{0.130}{\milli\second}$]
{
\includegraphics[width=0.31\textwidth]{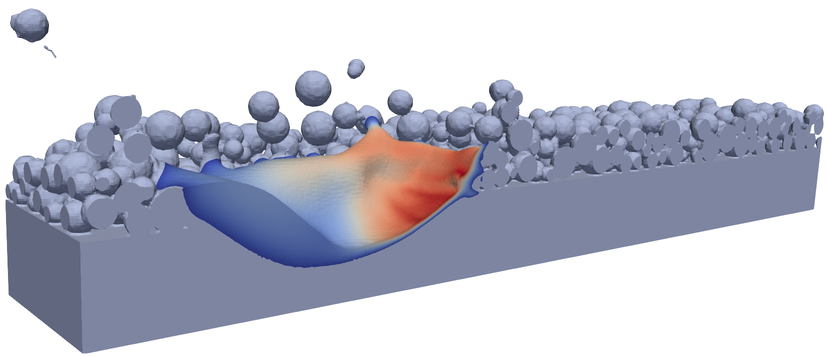}
}
\subfigure [$t = \SI{0.210}{\milli\second}$]
{
\includegraphics[width=0.31\textwidth]{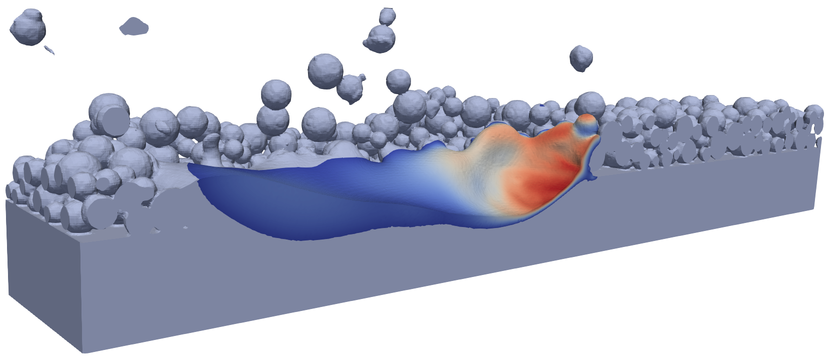}
}
\subfigure [$t = \SI{0.290}{\milli\second}$]
{
\includegraphics[width=0.31\textwidth]{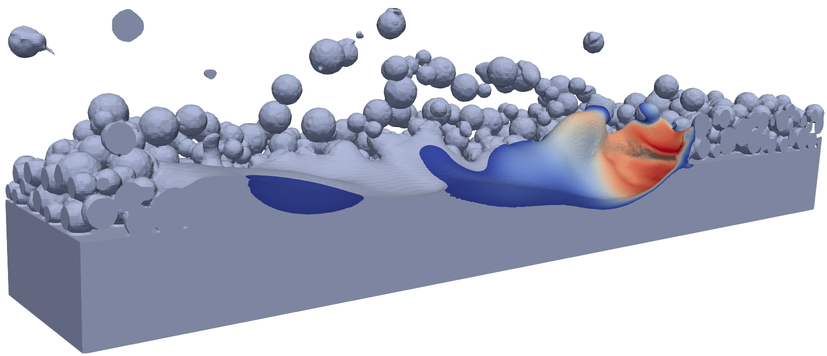}
}
\subfigure [$t = \SI{0.370}{\milli\second}$]
{
\includegraphics[width=0.31\textwidth]{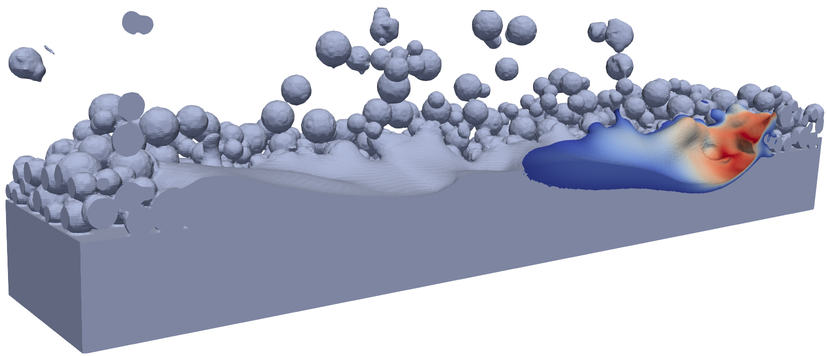}
}
\subfigure [$t = \SI{0.410}{\milli\second}$]
{
\includegraphics[width=0.31\textwidth]{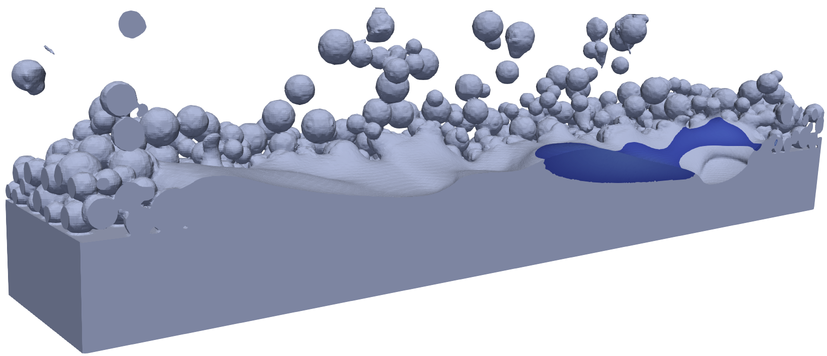}
}
\subfigure [$t = \SI{0.450}{\milli\second}$]
{
\includegraphics[width=0.31\textwidth]{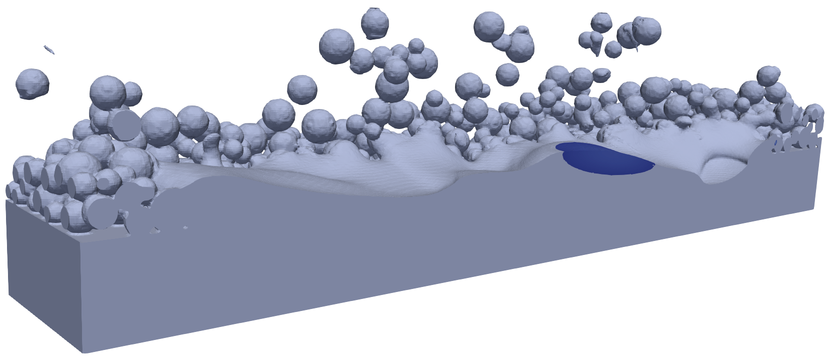}
}
\subfigure [$t = \SI{0.490}{\milli\second}$]
{
\includegraphics[width=0.31\textwidth]{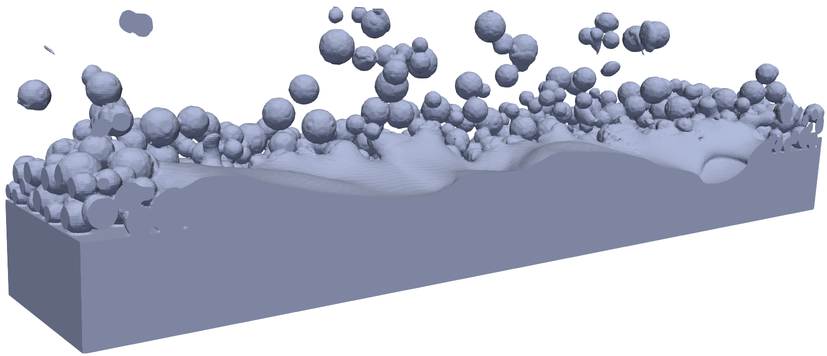}
}
\caption{Sectional view of line melting PBF example: time series illustrating the powder dynamics and the melt pool shape with temperature field ranging from $\SI{1700}{\kelvin}$ (blue) to $\SI{3400}{\kelvin}$ (red).}
\label{fig:pbf_linemelting_half}
\end{figure}

\begin{figure}[htbp]
\centering
\subfigure [$t = \SI{0.000}{\milli\second}$]
{
\includegraphics[width=0.31\textwidth]{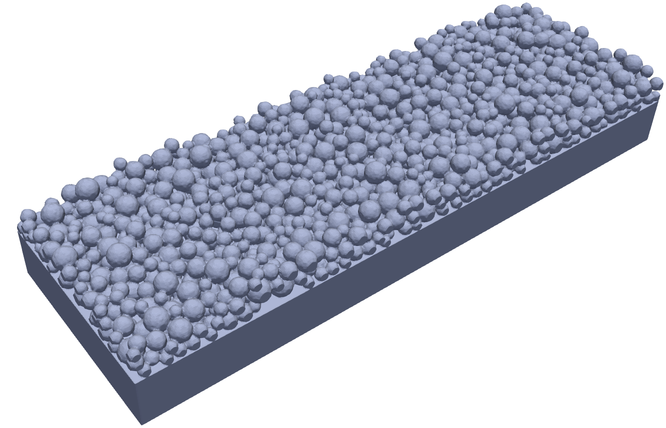}
}
\subfigure [$t = \SI{0.050}{\milli\second}$]
{
\includegraphics[width=0.31\textwidth]{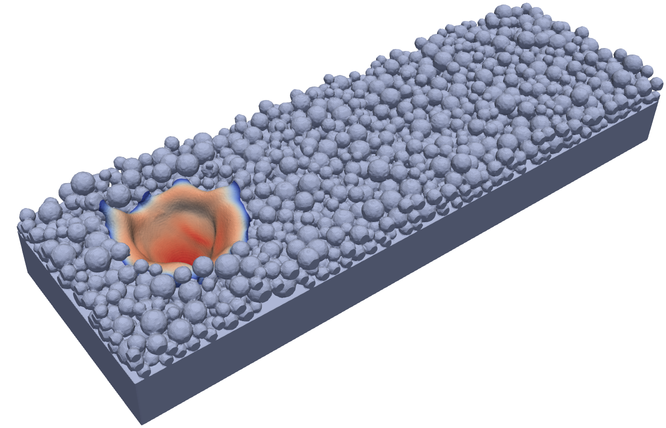}
}
\subfigure [$t = \SI{0.130}{\milli\second}$]
{
\includegraphics[width=0.31\textwidth]{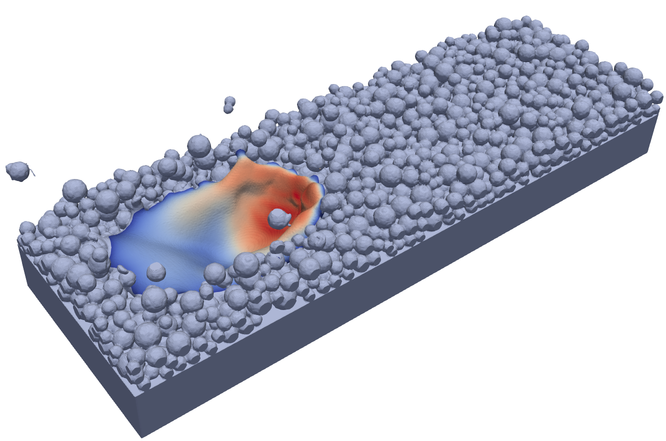}
}
\subfigure [$t = \SI{0.210}{\milli\second}$]
{
\includegraphics[width=0.31\textwidth]{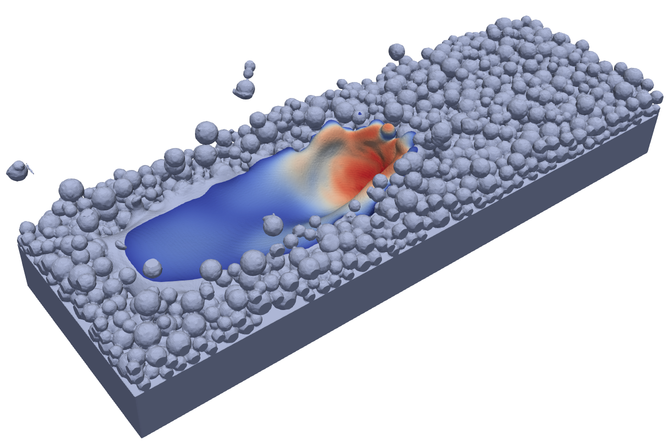}
}
\subfigure [$t = \SI{0.290}{\milli\second}$]
{
\includegraphics[width=0.31\textwidth]{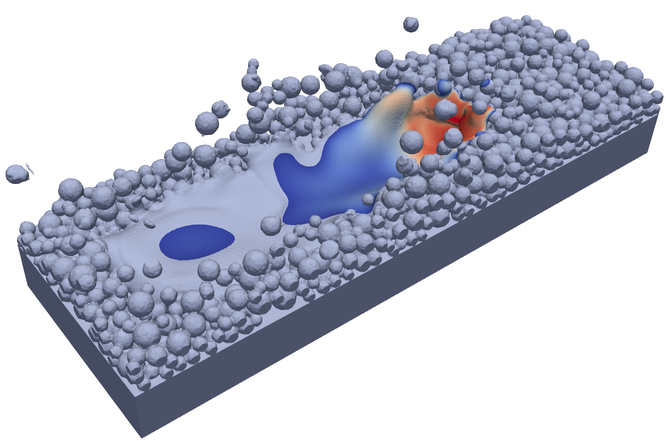}
}
\subfigure [$t = \SI{0.370}{\milli\second}$]
{
\includegraphics[width=0.31\textwidth]{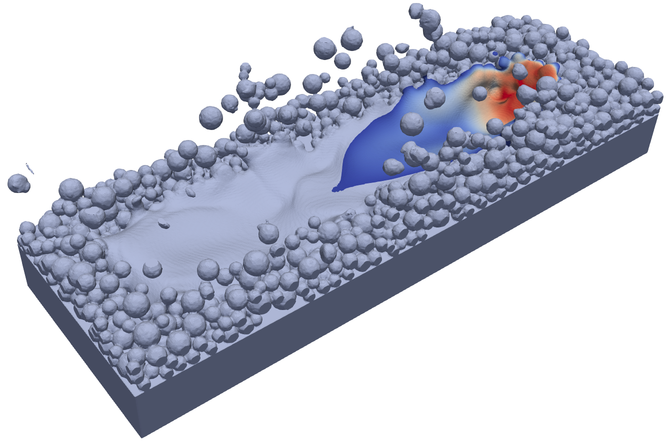}
}
\subfigure [$t = \SI{0.410}{\milli\second}$]
{
\includegraphics[width=0.31\textwidth]{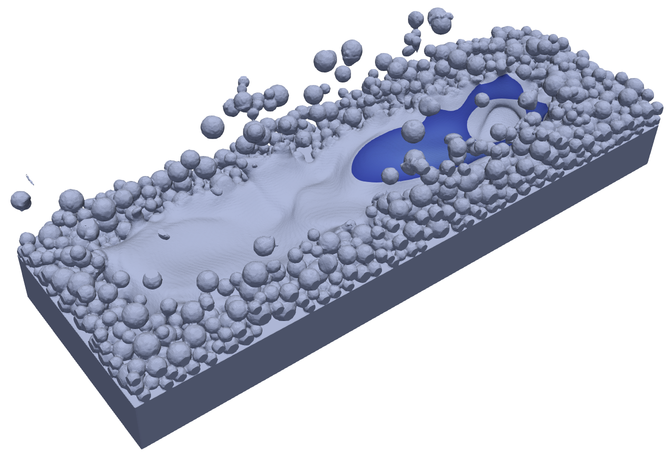}
}
\subfigure [$t = \SI{0.450}{\milli\second}$]
{
\includegraphics[width=0.31\textwidth]{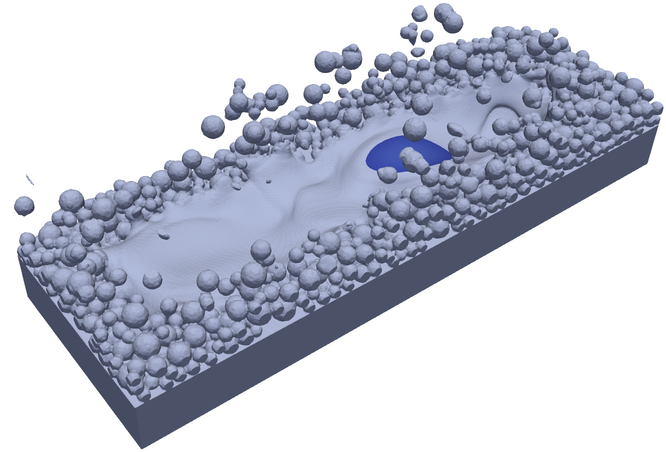}
}
\subfigure [$t = \SI{0.490}{\milli\second}$]
{
\includegraphics[width=0.31\textwidth]{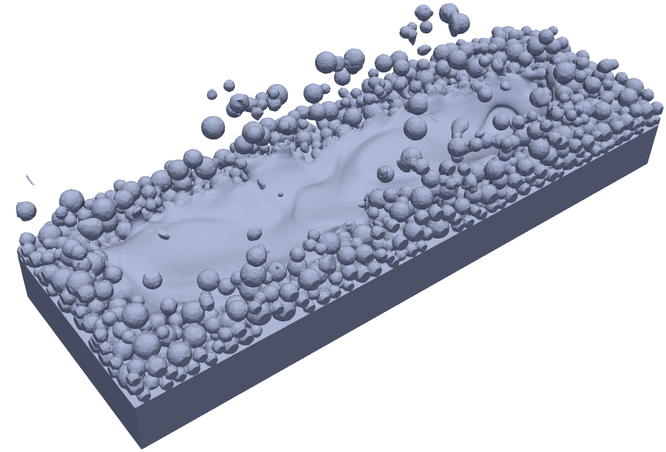}
}
\caption{Overall view of line melting PBF example: time series illustrating the powder dynamics and the melt pool shape with temperature field ranging from $\SI{1700}{\kelvin}$ (blue) to $\SI{3400}{\kelvin}$ (red).}
\label{fig:pbf_linemelting_full}
\end{figure}

\begin{figure}[htbp]
\centering
\subfigure [$t = \SI{0.000}{\milli\second}$]
{
\includegraphics[width=0.46\textwidth]{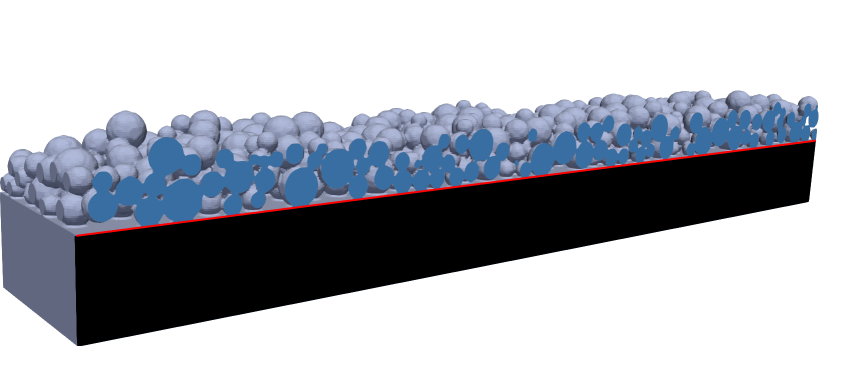}
\includegraphics[width=0.46\textwidth]{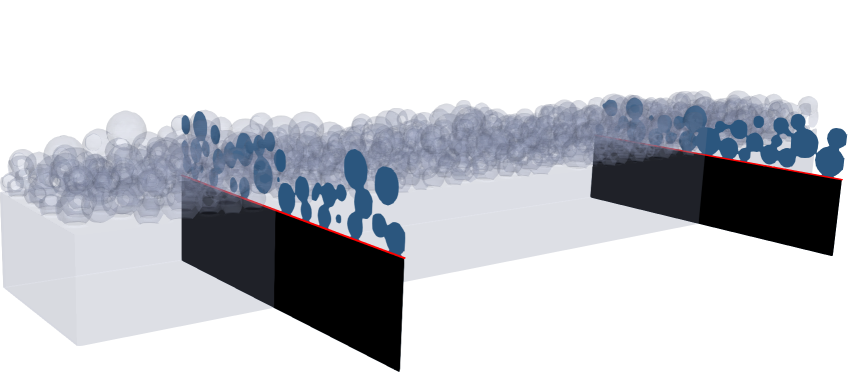}
}
\subfigure [$t = \SI{0.370}{\milli\second}$]
{
\includegraphics[width=0.46\textwidth]{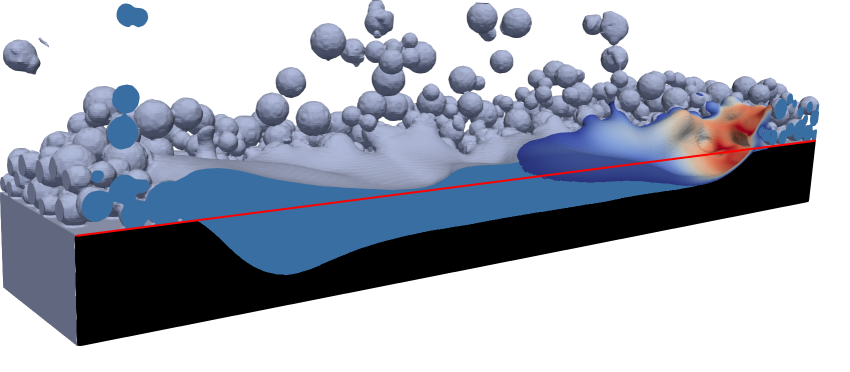}
\includegraphics[width=0.46\textwidth]{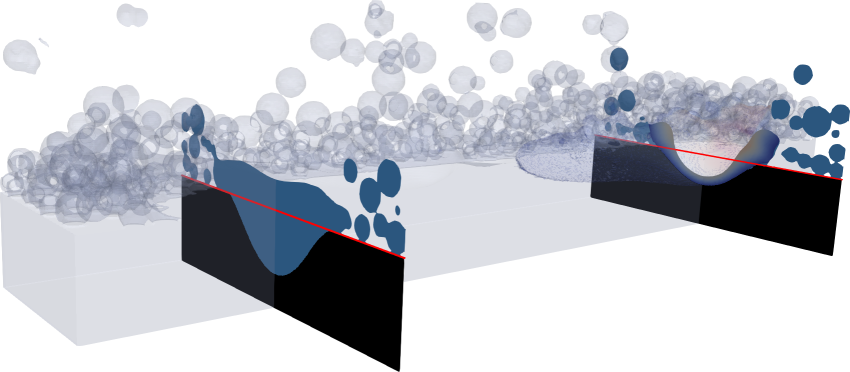}
}
\subfigure [$t = \SI{0.490}{\milli\second}$]
{
\includegraphics[width=0.46\textwidth]{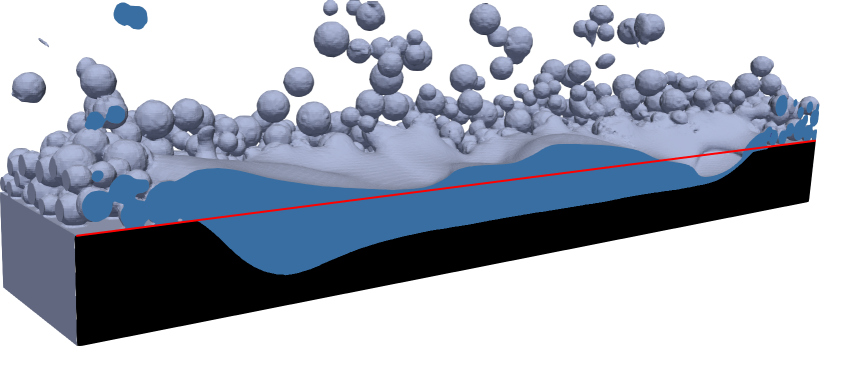}
\includegraphics[width=0.46\textwidth]{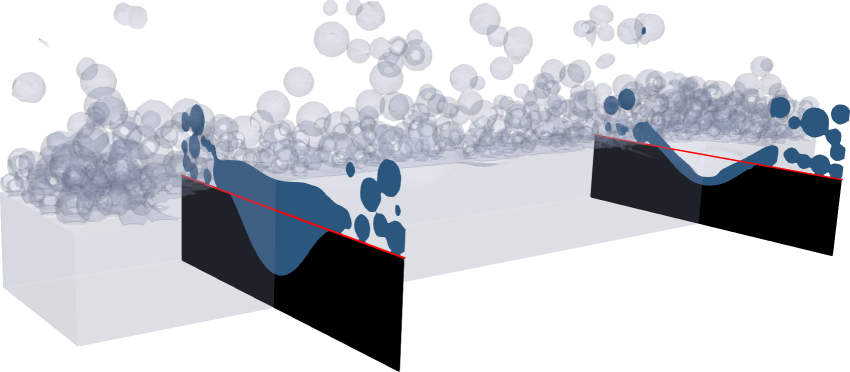}
}
\caption{Detail view of line melting PBF example: time series illustrating the powder dynamics and the melt pool shape with temperature field ranging from $\SI{1700}{\kelvin}$ (blue) to $\SI{3400}{\kelvin}$ (red) and sectional planes with non-molten material (black), resolidified material (blue), and surface of initial substrate (red).}
\label{fig:pbf_linemelting_detail}
\end{figure}

Finally, Figure~\ref{fig:pbf_linemelting_detail} gives a detail view of the line melting PBF example illustrating the non-molten and resolidified portion of the material. At the initial position of the laser, both the melt pool depth and the surface height of the solidified material are largest. In contrast, at the final laser position, the melt pool depth is smallest and, due to the remaining surface indentation after laser shut-off, the surface profile of the solidified track even lies below the initial height of the substrate. All together, the surface profile along the entire melt track is very wavy, which can be attributed to surface tension effects (e.g., Plateau-Rayleigh instabilities) and dynamic fluctuations of the melt pool shape due to coupled melt-powder-recoil pressure interactions.

Altogether, both variants of this example, i.e., point and line melting, demonstrate that the proposed modeling framework can robustly capture highly dynamic laser-melt-powder interactions while resolving important physical phenomena such as recoil pressure-induced wave patterns on the keyhole walls with a high detail. Finally, it shall be noted, that currently no frictional contact forces between powder particles are considered, and that the vapor jet and thereby induced gas flow are not explicitly modeled. These aspects will be in the focus of the authors' future research work. However, even when considering these additional effects, the general mechanisms observed in the studies above are expected to be still valid. Moreover, the general suitability of the proposed modeling framework for complex PBF processes is shown such that it can be recommended for further detailed studies.

\section{Conclusion and outlook} \label{sec:concl}

In this work, a general SPH modeling framework for coupled microfluid-powder dynamics problems involving thermo-capillary flow and reversible phase transitions was presented. Herein, parallel implementation aspects were not discussed. However, in the authors' previous work~\cite{Fuchs2021b}, a concept for the parallelization of the computational framework for fluid-solid and contact interaction problems was proposed and verified on the basis of an extensive scalability study. This framework enabled large-scale simulations such as the powder bed fusion line melting example, which was discretized by approximately \num{17.3e6} SPH particles and \num{5e5} time steps. Note that the implementation of such a parallel computational framework is far from trivial but indispensable when examining three-dimensional application-motivated examples that are of practical relevance.

The proposed modeling framework is suitable for the simulation of complex AM processes such as binder jetting (BJT), material jetting (MJT), directed energy deposition (DED), and powder bed fusion (PBF). To this end, the generality and robustness of the proposed computational modeling framework were demonstrated by examining several three-dimensional application-motivated examples. The obtained numerical results showcase the model's general ability to capture relevant physical phenomena such as coupled microfluid-powder dynamics and thermo-hydrodynamics involving surface tension and wetting effects as well as reversible phase transitions. In particular, the following key characteristics of the considered AM processes were observed:

\begin{itemize}
\item The dynamic impact of binder droplets in BJT evokes significant distortions of the powder packing structure, which can extend to droplet splashing and powder particle ejection in case of high droplet impact velocities.
\item Based on the preheating temperature of the substrate and the timing of successive droplets in MJT, aspects such as remelting depth or the occurrence of droplet coalescence and surface ripples can be controlled.
\item Early laser shut-off, or equivalently insufficient laser powers, in DED can lead to inclusions of partially molten powder particles.
\item As consequence of the evaporation-induced recoil pressure forces in PBF, powder particles are dynamically ejected and the powder feedstock is pushed ahead by the propagating melt pool.
\end{itemize}

In summary, it can be stated that the proposed modeling framework has the ability to accurately model important physical phenomena of a host of complex AM processes, and thus can be expected to become a valuable tool for detailed studies in this field. Most of these effects could only be represented due to the coupled microfluid-powder dynamics, considered for the first time in the proposed modeling framework. This modeling framework could also be used to model aspects of other AM processes than the ones considered, e.g., material extrusion or vat photopolymerization.

\section*{Acknowledgments}

This work was supported by funding of the Deutsche Forschungsgemeinschaft (DFG, German Research Foundation) within project
437616465 and project 414180263.

\section*{Declaration of Competing Interests}

The authors declare that they have no known competing financial interests or personal relationships that could
have appeared to influence the work reported in this paper.

\appendix
\section{Supplementary data}

Supplementary data related to the numerical examples of this work is available online in the corresponding ancillary files section of arXiv.

\begin{itemize}
\item \textbf{Videos 1 and 2:} Sectional and overall view of BJT example with initial velocity $\SI{1}{\meter\per\second}$ of the liquid binder droplet until $t = \SI{0.100}{\milli\second}$.
\item \textbf{Videos 3 and 4:} Sectional and overall view of BJT example with initial velocity $\SI{10}{\meter\per\second}$ of the liquid binder droplet until $t = \SI{0.100}{\milli\second}$.
\item \textbf{Videos 5 and 6:} Sectional and overall view of BJT example with initial velocity $\SI{50}{\meter\per\second}$ of the liquid binder droplet until $t = \SI{0.100}{\milli\second}$.
\item \textbf{Videos 7 and 8:} Sectional and overall view of MJT example with one molten metal droplet and initial temperature $\SI{500}{\kelvin}$ of the substrate until $t = \SI{0.125}{\milli\second}$ and temperature field ranging from $\SI{1700}{\kelvin}$~(blue) to $\SI{2500}{\kelvin}$~(red).
\item \textbf{Videos 9 and 10:} Sectional and overall view of MJT example with one molten metal droplet and initial temperature $\SI{1500}{\kelvin}$ of the substrate until $t = \SI{0.175}{\milli\second}$ and temperature field ranging from $\SI{1700}{\kelvin}$~(blue) to $\SI{2500}{\kelvin}$~(red).
\item \textbf{Video 11:} Overall view of MJT example with two molten metal droplets and initial temperature $\SI{500}{\kelvin}$ of the substrate until $t = \SI{0.200}{\milli\second}$ and temperature field ranging from $\SI{1700}{\kelvin}$~(blue) to $\SI{2500}{\kelvin}$~(red).
\item \textbf{Video 12:} Overall view of MJT example with three molten metal droplets and initial temperature $\SI{500}{\kelvin}$ of the substrate until $t = \SI{0.200}{\milli\second}$ and temperature field ranging from $\SI{1700}{\kelvin}$~(blue) to $\SI{2500}{\kelvin}$~(red).
\item \textbf{Video 13:} Overall view of DED example with laser continuously switched on until $t = \SI{0.020}{\milli\second}$ and temperature field ranging from $\SI{1700}{\kelvin}$~(blue) to $\SI{3400}{\kelvin}$~(red).
\item \textbf{Video 14:} Overall view of DED example with laser switched off at $t = \SI{0.012}{\milli\second}$ until $t = \SI{0.035}{\milli\second}$ and temperature field ranging from $\SI{1700}{\kelvin}$~(blue) to $\SI{3400}{\kelvin}$~(red).
\item \textbf{Videos 15 and 16:} Sectional and overall view of point melting PBF example until $t = \SI{0.250}{\milli\second}$ and temperature field ranging from $\SI{1700}{\kelvin}$~(blue) to $\SI{3400}{\kelvin}$~(red).
\item \textbf{Videos 17 and 18:} Sectional and overall view of line melting PBF example until $t = \SI{0.490}{\milli\second}$ and temperature field ranging from $\SI{1700}{\kelvin}$~(blue) to $\SI{3400}{\kelvin}$~(red).
\end{itemize}

\section{Representative material and discretization parameters}

This appendix provides representative material parameters for molten metal, solid metal, atmospheric gas, and liquid binder, given in the Tables~\ref{tab:material_params_molten_metal}, \ref{tab:material_params_solid_metal}, \ref{tab:material_params_atmoshperic_gas}, and \ref{tab:material_params_liquid_binder}, as applied in the numerical examples in this work. Note that surface tension related parameters are given in the Tables for the liquid phase, i.e., molten metal or liquid binder, against atmospheric gas. Besides, this appendix provides the applied discretization parameters given in Table~\ref{tab:discretization_params}.

\begin{table}[htbp]
    \centering
    \caption{Representative material parameters for molten metal (stainless steel).}
    \label{tab:material_params_molten_metal}
    \begin{tabular}{p{1.5cm} p{8.0cm} p{3.0cm} p{2.0cm}}
        \toprule
        Symbol & Property & Value & Units \\
        \midrule
        $\rho_0$ & Reference density& \num{7430} & \si{\kilogram\per\meter\cubed} \\
        $\eta$ & Dynamic viscosity& \num{6.0e-3} & \si{\kilogram\per\meter\per\second} \\
        $\alpha_0$ & Surface tension coefficient at reference temperature & \num{1.8} & \si{\newton\per\meter}  \\
        $\alpha_{\min}$ & Minimum surface tension & \num{0.2} & \si{\newton\per\meter} \\
        $T_{\alpha_0}$ & Reference temperature for surface tension& \num{1700} & \si{\kelvin} \\ 
        $\alpha'_0$ & Surface tension gradient coefficient& \num{-1.0e-3} & \si{\newton\per\meter\per\kelvin} \\
        $\theta_0$ & Equilibrium wetting angle & \num{60} & \si{\degree} \\ 
        $T_m$ & Melt temperature & \num{1700} & \si{\kelvin}  \\
        $\Delta{}T_{s}$ & Surface tension regularization temperature interval & \num{5.0} & \si{\kelvin} \\
        $\Delta{}T_{d}$ & Viscous dissipation force regularization temperature interval & \num{300} & \si{\kelvin} \\
        $T_v$ & Boiling temperature & \num{3000} & \si{\kelvin}  \\ 
        $c_p$ & Heat capacity & \num{965} & \si{\joule\per\kilogram\per\kelvin} \\ 
        $k$ & Thermal conductivity & \num{35.95} & \si{\watt\per\meter\per\kelvin} \\ 
        $\zeta_l$ & Laser absorptivity & \num{0.5} & $ - $  \\ 
        $C_P$ & Pressure constant of recoil pressure model & \num{5.4e4} & \si{\newton\per\meter\squared} \\ 
        $C_T$ & Temperature constant of recoil pressure model & \num{5e4} & \si{\kelvin}  \\ 
        $h_v$ & Latent heat of evaporation & \num{6.583e6} & \si{\joule\per\kilogram} \\ 
        $T_{h,0}$ & Reference temperature for specific enthalpy & \num{663.731} & \si{\kelvin}  \\ 
        $C_M$ & Constant for vapor mass flow & \num{1.1095e-3} & \si{\kelvin\second\squared\per\meter\squared} \\
        $\alpha^{lg}_{0}$ & Artificial viscosity factor liquid-gas & \num{7.2} & $ - $ \\
        $\alpha^{sf}_{0}$ & Artificial viscosity factor solid-fluid & \num{1.0} & $ - $ \\
        \bottomrule
    \end{tabular}
\end{table}

\begin{table}[htbp]
    \centering
    \caption{Representative material parameters for solid metal (stainless steel).}
    \label{tab:material_params_solid_metal}
    \begin{tabular}{p{1.5cm} p{8.0cm} p{3.0cm} p{2.0cm}}
        \toprule
        Symbol & Property & Value & Units \\
        \midrule
        $\rho_0$ & Reference density& \num{7430} & \si{\kilogram\per\meter\cubed} \\
        $c_p$ & Heat capacity & \num{965} & \si{\joule\per\kilogram\per\kelvin} \\ 
        $k$ & Thermal conductivity & \num{35.95} & \si{\watt\per\meter\per\kelvin} \\
        $\zeta_l$ & Laser absorptivity & \num{0.5} & $ - $  \\
        \bottomrule
    \end{tabular}
\end{table}

\begin{table}[htbp]
    \centering
    \caption{Representative material parameters for atmospheric gas.}
    \label{tab:material_params_atmoshperic_gas}
    \begin{tabular}{p{1.5cm} p{8.0cm} p{3.0cm} p{2.0cm}}
        \toprule
        Symbol & Property & Value & Units \\
        \midrule
        $\rho_0$ & Reference density& \num{7.43} & \si{\kilogram\per\meter\cubed} \\
        $\eta$ & Dynamic viscosity& \num{6.0e-4} & \si{\kilogram\per\meter\per\second} \\
        $c_p$ & Heat capacity & \num{10.0} & \si{\joule\per\kilogram\per\kelvin} \\
        $k$ & Thermal conductivity & \num{0.026} & \si{\watt\per\meter\per\kelvin} \\
        $\zeta_l$ & Laser absorptivity & \num{0.0} & $ - $  \\ 
        \bottomrule
    \end{tabular}
\end{table}

\begin{table}[htbp]
    \centering
    \caption{Representative material parameters for liquid binder.}
    \label{tab:material_params_liquid_binder}
    \begin{tabular}{p{1.5cm} p{8.0cm} p{3.0cm} p{2.0cm}}
        \toprule
        Symbol & Property & Value & Units \\
        \midrule
        $\rho_0$ & Reference density& \num{1000} & \si{\kilogram\per\meter\cubed} \\
        $\eta$ & Dynamic viscosity& \num{1.0e-3} & \si{\kilogram\per\meter\per\second} \\
        $\alpha_0$ & Surface tension coefficient & \num{0.5} & \si{\newton\per\meter}  \\
        $\theta_0$ & Equilibrium wetting angle & \num{60} & \si{\degree} \\
        $\alpha^{lg}_{0}$ & Artificial viscosity factor liquid-gas & \num{1.8} & $ - $ \\
        \bottomrule
    \end{tabular}
\end{table}

\begin{table}[htbp]
    \centering
    \caption{Discretization parameters}
    \label{tab:discretization_params}
    \begin{tabular}{p{1.5cm} p{8.0cm} p{3.0cm} p{2.0cm}}
        \toprule
        Symbol & Property & Value & Units \\
        \midrule
        $W$ & Smoothing kernel & Quintic spline~\cite{Morris1997} & \\
        h & Smoothing length & $1.\overline{6}$ & \si{\micro\meter} \\
        $r_c$ & Support radius & $5.0$ & \si{\micro\meter} \\
        $\Delta{}x$ & Initial particle spacing & $1.\overline{6}$ & \si{\micro\meter} \\
        $\Delta{}t$ & Time step size & \num{1.0e-6} & \si{\milli\second} \\
        \bottomrule
    \end{tabular}
\end{table}

\bibliographystyle{elsarticle-num} 
\bibliography{collection.bib}

\end{document}